\documentclass[double]{article}
\usepackage{times}
\usepackage{algorithm}
\usepackage{algorithmic}
\usepackage{wrapfig}
\usepackage{verbatim}
\usepackage{amsmath}
\usepackage{graphicx}
\usepackage{subcaption}
\usepackage[active]{srcltx}
\usepackage{color}
\usepackage{lineno}
\usepackage{dirtytalk}
\usepackage{amsthm}
\usepackage{authblk}
\usepackage{listings}
\usepackage{setspace} 
\usepackage{tikz}
\usetikzlibrary{shapes.geometric, arrows}

\usepackage{epsfig,graphicx,amsfonts}
\usepackage{amsmath,amsthm,amssymb}
\usepackage{xcolor}

\usepackage{adjustbox}
\usepackage{multirow}
\usepackage{setspace}
\usepackage{hyperref}
\usepackage{comment}

\long\def\comment #1\commentend{}

\begin{document}

\title{\Large An Empirically-parametrized Spatio-Temporal Extended-SIR Model for Combined Dilution and Vaccination Mitigation for Rabies Outbreaks in Wild Jackals}

\author{Teddy Lazebnik$^{1,*}$, Yehuda Samuel$^{2}$, Jonathan Tichon$^{2}$, Roi Lapid$^{3}$, Roni King$^{4}$, Tomer Nissimian$^{3,4}$, Orr Spiegel$^{2}$\\
\(^1\) Department of Cancer Biology, Cancer Institute, University College London, London, UK\\
\(^2\) School of Zoology, Faculty of Life Sciences, Tel Aviv University, Tel Aviv, Israel \\
\(^3\) Koret School of Veterinary Medicine, Faculty of Agriculture, Food and Environment, The Hebrew University of Jerusalem, Rehovot, Israel\\
\(^4\) Israel Nature and Parks Authority \\
\(^*\) Corresponding author: lazebnik.teddy@gmail.com \\
}

\date{ }

\maketitle 

\begin{abstract}
\noindent
The transmission of zoonotic diseases between animals and humans poses an increasing threat. Rabies is a prominent example with various instances globally, facilitated by a surplus of meso-predators (commonly, facultative synanthropic species e.g., golden jackals [\textit{Canis aureus}, hereafter jackals]) thanks to the abundance of anthropogenic resources leading to dense populations close to human establishments. To mitigate rabies outbreaks and prevent human infections, authorities target the jackal which is the main rabies vector in many regions, through the dissemination of oral vaccines in known jackals' activity centers, as well as opportunistic culling to reduce population density. Because dilution (i.e., culling) is not selective towards sick or un-vaccinated individuals, these two complementary epizootic intervention policies (EIPs) can interfere with each other. Nonetheless, there is only limited examination of the interactive effectiveness of these EIPs and their potential influence on rabies epizootic spread dynamics, highlighting the need to understand these measures and the spread of rabies in wild jackals. In this study, we introduce a novel spatio-temporal extended-SIR (susceptible-infected-recovered) model with a graph-based spatial framework for evaluating mitigation efficiency. We implement the model in a case study using a jackal population in northern Israel, and using spatial and movement data collected by Advanced Tracking and Localization of Animals in real-life Systems (ATLAS) telemetry. An agent-based simulation approach allows us to explore various biologically-realistic scenarios, and assess the impact of different EIPs configurations. Our model suggests that under biologically-realistic underlying assumptions and scenarios, the effectiveness of both EIPs is not influenced much by the jackal population size but is sensitive to their dispersal between activity centers. Furthermore, we show both theoretically and empirically that counter-intuitively, there are cases in which under the practice of EIPs, the epizootic (or endemic) spread is increasing (due to interference among them). Our findings emphasize the importance of accurately capturing the local jackal movement dynamics to obtain and predict the desired outcome from an applied EIP configuration. \\ \\

\end{abstract}

\maketitle \thispagestyle{empty}
\pagestyle{myheadings} \markboth{Draft:  \today}{Draft:  \today}
\setcounter{page}{1}

\section{Introduction}
\label{sec:introduction}
Zoonotic diseases are a growing threat globally, responsible for over 15\% of human annual mortality \cite{orr_1}. Increasing prevalence reflects the interactive effects of land-use changes, shifts in community assembly, and the abundance and behavior of various wildlife species  \cite{orr_2,intro_2,orr_3}. The OneHealth approach highlights the dependencies and feedback among humans, livestock, and wildlife, and that for mitigating disease spread we need to investigate their interactions and growing friction \cite{con_new_5}. While most species are negatively impacted by the ongoing anthropogenic development (i.e. showing population reductions or extinctions), a small subset is becoming synanthropic, benefiting from human resources (e.g. waste) and habitat modifications. These species often become overabundant inside or near human settlements, playing a unique role in epidemic outbreaks and potential for zoonotic disease, with examples like feral pigeons and Salmonella or canids and Rabies. Because of their high densities and mobility, these species may pose a threat to ecosystems and humans alike \cite{orr_4}. Indeed, the interconnection between disease ecology and animal movement ecology is well established \cite{intro_1,intro_0,intro_2}. On the one hand, hosts' movement and behavior show high diversity across environmental and ecological conditions, affecting the probabilities of both encountering a pathogen and spreading it. On the other hand, host movement and social behaviors are also affected by pathogen presence, either due to sickness, or due to intentional host manipulation by the pathogen \cite{intro_1,con_new_3}. Rabies, for instance, may induce aggressive behavior in its mammalian hosts facilitating direct transmission through bites of susceptible subjects. Thus, these interfaces of movement and pathogens are connected both directly and indirectly to the general spread of epizootic, and highlight the importance of modeling disease spread in an ecological context considering such dependencies for predicting the effectiveness of different interventions \cite{intro_3}. 

Mathematical models have been shown to be useful tools for studying disease ecology and outbreak dynamics \cite{lv_pandemic_3,lv_pandemic_4,teddy_multi_strain,multi_populations_1,multi_populations_4}. These models can be divided into two main groups - statistical models and compartmental models. The statistical models use signal-processing and machine learning based models to predict properties of the epizootic over time \cite{multi_populations_3,first_teddy_paper,different_approach_from_sir,different_approach_from_sir_3,different_approach_from_sir_2}. While these models show promising results in predicting epidemiological properties, they are extremely sensitive to the epizootic properties and environmental condition. Further, these models often operate as \say{black boxes} which does not allow for investigation of the underlying dynamics in an epizootic, limiting the validity of these predictions across changing conditions \cite{intro_models_1}. Compartmental models, in contrast, are based on a state-machine of epidemiological states that the individuals in the population transform between; but may fail to capture dependencies between states and behaviors or heterogeneity within a compartment (e.g. among individuals' behavior) \cite{orr_r_1,teddy_pandemic_management}. A significant portion of the compartmental models are based on the SIR (susceptible-infected-recovered) epidemiological model \cite{first_sir} which assumes three epidemiological states - a susceptible state where individuals are healthy and can be infected, an infected state where individuals are infected, and can also infect susceptible individuals, and the recovered state where individuals are no longer infected and have immunity. The compartmental models extend the SIR model by introducing both temporal and spatial components such as two-age groups \cite{first_teddy_paper}, the exposed state \cite{bio_2}, population movement \cite{spatial_2_example}, and multi-strain \cite{multi_strain_3}. 

One of the strengths of these basic and advanced SIR models is the ability to predict the effectiveness of alternative epizootic intervention policies (EIPs) in general contexts \cite{pip_5,pip_2,pip_3}. However, many of the models focus on one specific EIP, and consideration of the combined effects of alternative EIPs is less common \cite{pip_4,pip_1}. Several studies focusing on human pathogens demonstrated that EIPs may have an interactive effect. For instance, \cite{teddy_pandemic_management} working on COVID-19 dynamics demonstrated that reducing working and schooling hours to reduce spread exposure can interfere with the efficiency of closing large social events if the first EIP is aggressively applied while the latter is not. Although such studies underscore the importance of modeling the interactive effectiveness of more than one EIP, and studies in animal ecology highlight potential feedback in disease dynamics, to date, models considering combined EIPs in an ecological context are absent \cite{base_paper}. In this study, we aim to address this knowledge gap by focusing on a free-ranging population of golden jackals (\textit{Canis aureus}, hereafter simply jackals) that resides in North-Eastern Israel, and subsidized by anthropogenic resource. Nowadays, the jackal is the main wildlife vector for rabies outbreaks in Israel \cite{rabis_israel}, and hence the importance of investigating the effectiveness and potential interference among commonly applied EIPs in this population. 

Rabies is one of the earliest documented ailments and retains its status as the foremost viral zoonotic disease globally \cite{rabis_bad}. Currently, more than 59,000 people die annually from rabies around the world, accruing an estimated cost of more than \$8.6 billion per year \footnote{For the full report: \url{https://www.who.int/news-room/fact-sheets/detail/rabies/?gad_source=1&gclid=Cj0KCQjwsuSzBhCLARIsAIcdLm4MM080pShuTZfRUO2GRgkrstfvVxOHSr2qhy3Xk8iwOu3x31y-p3oaAqOaEALw_wcB}}, and recent studies emphasize the OneHealth approach relevance for the disease, with human death associated with perturbation of natural ecosystems \cite{orr_r_2}. Typically, the rabies virus (\textit{Rabies Iyssavirus}) spreads primarily through bites from already infected animals \cite{rabis_bite_spread}. While diverse mammals are susceptible to the virus, canid species tend to be the prominent hosts in most cases (and occasionally raccoons [\textit{Procyon lotor}] and striped skunks [\textit{Mephitis mephitis}]), with examples ranging from stray dogs (\textit{Canis familaris}) in India to jackals species (\textit{Canis adustus} and \textit{Canis mesomelas}) in Zimbabwe \cite{rabis_jackle_1,rabis_jackle_2}. In wildlife reservoirs, various rabies virus variants have been discovered. These variants are geographically separated and genetically distinct, each associated with different terrestrial hosts including raccoons, skunks, foxes, coyotes, and bats \cite{rabits_g_1,rabits_g_2,rabits_g_3}. For instance, in Israel, red fox [\textit{Vulpes vulpes}] was previously the main host, until a spread of a new strain, mostly transmitted by jackals, emerged. Following a bite, the virus migrates from the wound through the peripheral nervous system into the central nervous system and ultimately the brain \cite{rabis_inside_host}. The duration between the initial infection and the onset of symptoms varies somewhat among and within species and may hinge on factors such as the bite location and severity of the wound. However, afflicted animals typically succumb to the disease within a week of symptom onset \cite{rabis_israel}. Post-exposure vaccinations are effective for humans, but typically, once the virus reaches the brain and symptoms appear the disease is fatal, highlighting the importance of prevention for this particular disease \cite{rabis_fatal}.

Preventive measures for rabies outbreaks are challenged by the need to reduce prevalence and transmission rates in free-ranging wildlife populations \cite{rabis_bite_spread}. Management often relies on one or two main EIPs: oral vaccination and culling (hereafter \say{population dilution}) \cite{rabits_g_2}. Vaccination relies on the spreading of either modified-live, highly attenuated, or recombinant viruses contained within attractive, edible baits. Vaccine-laden baits are distributed strategically by hand, at baiting stations, or via aircraft \cite{Rupprecht2024}. Beyond direct removal of symptomatic individuals, it is typically impossible to target infected (but asymptomatic) or non-vaccinated during culling efforts aimed at reducing the host population's density. While the efficiency of vaccination is well-established, the efficiency of the latter is more context-dependent \cite{morters_control_rabies_2013}, and the possible interdependency between the two is largely unknown. We hypothesize that in certain biologically relevant conditions, non-selective population dilution can achieve a negative impact by removing vaccinated individuals or by enhancing host movement and facilitating transmission. In this study, we aim to address this knowledge gap by developing an extended SIR model implemented with agent-based simulations for a specific study system. The simulations are parameterized with empirical field data complemented with parameters based on published literature and expert opinion. This serves as a preliminary step to outline the methodology and validate its feasibility.

In Israel (and neighboring countries), rabies is prevalent in natural and agricultural systems, with frequent outbreaks \cite{gdalevich2000rabies} of this epizootic (or endemic). In particular, the native golden jackal is considered a prominent vector for the virus in the recent two decades, with dozens of confirmed positive cases annually \cite{rabis_israel}. Jackals' centrality in rabies outbreaks in Israel results in occasional instances of aggressive interactions between jackals and humans, and consequently substantial human-wildlife conflict. Israel Nature and Parks Authority (INPA) is mitigating the prevalence of rabies in this dense population through the dissemination of oral vaccines and opportunistic culling of jackals. Because, currently the spatial distribution of the rabies vaccine relies on subjective estimation and expert judgment with respect to jackal density and movement, it is imperative to establish a structured framework to guide the distribution strategies of the two EIPs. Accordingly, here we mark the initial phase in developing such a framework, utilizing spatial optimal control techniques. The rest of the paper is organized as follows: Section \ref{sec:model} outlines the proposed epidemiological model as well as the EIPs formalizations, agent-based simulation implementation of the proposed model, EIP optimization procedure, and fitting procedure for empirical data. Section \ref{sec:theory} presents a theoretical analysis of the model, including proof that the solution of the proposed model exists and is unique and that EIPs can also increase the rabies spread. Section \ref{sec:experiments} describes the empirical data acquisition used for the real-world experiments of the proposed model (including jackal telemetry tracking, aimed at quantifying features of their space use) and the results of these \textit{in silico} experiments. Finally, in section \ref{sec:discussion} we discuss the ecological applications and outcomes of our results in the context of OneHealth and suggest future directions of investigation. 

\section{Methods and Materials}
\label{sec:model}
In this section, we first formally introduce the proposed disease spread model for any jackal population with rabies with two possible EIPs. Afterward, we implement the proposed general model for the case of the Harod Valley in Israel.

A jackal population, \(J\), as well as the environment the population occupies, \((V, \zeta)\), participate in the dynamics of rabies spread and EIP efficiency. We divided the jackal population of \(n\) individuals into \(c\) activity centers (or sub-populations, geographic regions supporting a fraction of the overall population and is at least as spatially big as the jackals' home range), denoted by \(\forall k \in [1, \dots, c]: v_k\). Formally, jackals can move between activity centers but associated with one at any given point in time (no floaters). In each activity center, \(v_k \in V\), the jackal sub-population on-site follows four classical natural demographic processes along time \(t\): first, new jackals are born at a rate corresponding the jackal's population size; second, jackal naturally die at a given rate; third, jackals are immigrating into the activity centers (from other activity centers); finally, jackals are emigrating from their current activity center to others (i.e. dispersal) which mathematically represented as edges (\(\zeta \subset V \times V\)) in a graph of locations which represents the environment. The jackal population grows given enough food to support the population. Realistically, each activity center has a carrying capacity limiting the jackal population size. According to the ideal free distribution \cite{Fretwell1970} jackals may migrate from their activity center to other activity centers with more (\textit{per-capita}) available food. Hence, each activity center can be defined by the amount of available food, \(F\), it has for the jackals, and it generates new food at a constant rate with some carrying capacity. The jackal sub-population living in the activity center consumes this food at a rate proportional to its size. Jackals die due to a lack of available food or as a result of natural death like old age or other non-rabies related reasons like road kill (that is a common mortality factor in our system).

In addition to these underlying demographical processes, the jackals can be infected by rabies that may spread within and among activity centers. Jackals are divided into three epidemiological classes; the susceptible (\(S\)), the exposed (\(E\)), and the infected (\(I\)); such that \(J = S + E + I\). The susceptible individuals in the population can be infected with rabies virus. Since rabies is lethal for the jackals, we assume that there are no recovered individuals. Exposed individuals are already infected by the virus but not yet infectious themselves, eventually becoming infected and infectious to others until their death, a few weeks after initial infection. We assume that all epidemiological classes have a natural death while only susceptible individuals are healthy enough to participate in reproduction (in addition, given the fast cycle from Exposure to death, successful breeding is not feasible). Furthermore, while all epidemiological classes can migrate, they may have different rates. On the one hand, dispersal rates of infected individuals might be reduced due to the physical burden required from such a process \cite{social_dist,sick_walk_2}. On the other hand, enforced social distancing or host manipulation by the virus (known to induce aggressive behavior) may enhance dispersal rates \cite{social_dist}. In a similar manner, each epidemiological class may have a different average consumption of food. Importantly, we decided not to include a recovery in the model as there is little evidence for a naturally formed immune class \cite{no_recover_class}. Fig. \ref{fig:scheme} provides a schematic view of the proposed model.

\begin{figure}[!ht]
    \centering
    \includegraphics[width=0.99\textwidth]{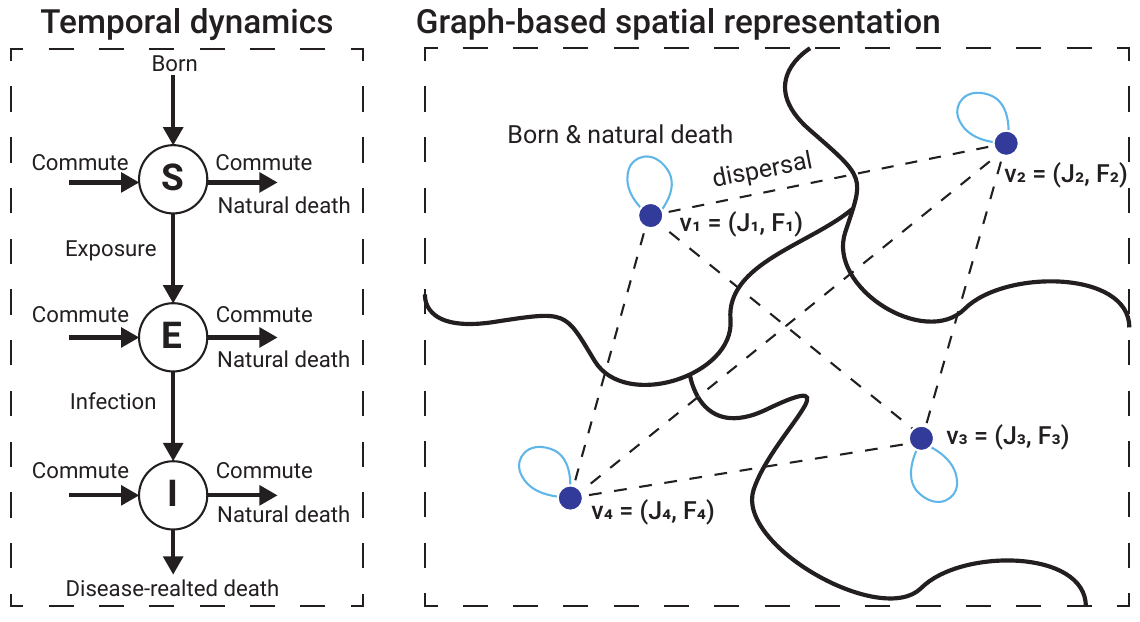}
    \caption{A schematic view of the proposed model. The left panel shows the temporal-epidemiological dynamics according to an SEI model. The right panel shows a graph-based spatial representation of the population with 4 activity centers, each with \(J_i\) jackals, and \(F_i\) food. Dashed lines represent emigration (dispersal) among centers.}
    \label{fig:scheme}
\end{figure}

\subsection{Model definition}
In this section, we mathematically formalize the model's dynamics while assuming the individuals in the jackal population to be identical. The mathematical formalization of many components is following the suggestions in \cite{orr_cool_paper}. For each activity center, \(v_k\), the jackal population size, divided into the susceptible, exposed, and infected epidemiological classes (\(S_k, E_k, I_k\)), and the amount of available food, \(F_k\), dynamics.

In Eq. (\ref{eq:s}), \(\frac{dS_k(t)}{dt}\) describes the dynamic amount of susceptible jackals in an activity center over time for the \(k_{th}\) activity center. Overall, this is affected by the following five terms, either contributing individuals (birth, immigration), or removing individuals (death, rabies progress, emigration). First, the susceptible population has a logarithmic growth with a rate proportional to the available food in the activity center (\(F_k(t)\)) with a carrying capacity \(\kappa_k\). Second, susceptible individuals are infected by infectious individuals at a rate \(\beta\). Third and fourth, the immigration (or dispersal) between activity centers \(i\) and \(j\) occurs at a rate \(m_{ij}^s\) and with respect to the relative difference in available \textit{per-capita} food. Finally, susceptible individuals naturally die at a rate \(\nu_k^s\).

\begin{equation} \begin{array}{l} \frac{dS_k(t)}{dt} = \underbrace{F_k(t) S_k(t) \big ( 1 - \frac{J_k(t)}{\kappa_k} \big )}_{\text{logarithmic growth}} - \underbrace{\beta S_k(t) I_k(t)}_{\text{infection}} \ + \underbrace{\sum_{j=1, j \neq k}^n \big ( m_{jk}^s \frac{S_j(t) \big ( F_k(t) - F_j(t) \big )}{J_k(t) F_k(t)} \big )}_{\text{immigration driven by food gradient}} \\ \\ - \underbrace{\sum_{j=1, j \neq k}^n \big ( m_{kj}^s \frac{S_k(t) \big ( F_k(t) - F_j(t) \big )}{J_j(t) F_k(t)} \big )}_{\text{emigration driven by food gradient}} \ - \underbrace{\nu_k^s S_k(t)}_{\text{natural death}}. \end{array} \label{eq:s} \end{equation}

In Eq. (\ref{eq:e}), \(\frac{dE_k(t)}{dt}\) describes the dynamic amount of currently exposed jackals in each activity center over time for the \(k_{th}\) activity center. 
Similarly to the susceptible individuals described above, this is also affected by the following five terms. First, susceptible individuals become exposed at a rate \(\beta\) and join the Exposed epidemiological state group. Second, exposed individuals are omitted from this as they transform into infected individuals at a rate \(\phi\). Third and fourth, the dispersal between activity centers \(i\) and \(j\) occurs at a rate \(m_{ij}^e\) and with respect to the relative difference in available food. Finally, exposed individuals naturally die (from non-rabies reasons) at a rate \(\nu_k^e\).

\begin{equation} \begin{array}{l} \frac{dE_k(t)}{dt} = \underbrace{\beta S_k(t) I_k(t)}_{\text{infection of susceptible individuals}} - \underbrace{\phi E_k(t)}_{\text{progression to infected state}} \ + \underbrace{\sum_{j=1, j \neq k}^n \big ( m_{jk}^e \frac{E_j(t) \big ( F_k(t) - F_j(t) \big )}{J_k(t) F_k(t)} \big )}_{\text{immigration driven by food gradient}} \\\\ - \underbrace{\sum_{j=1, j \neq k}^n \big ( m_{kj}^e \frac{E_k(t) \big ( F_k(t) - F_j(t) \big )}{J_j(t) F_k(t)} \big )}_{\text{emigration driven by food gradient}} \ - \underbrace{\nu_k^e E_k(t)}_{\text{natural death}}. \end{array} \label{eq:e} \end{equation}

In Eq. (\ref{eq:i}), \(\frac{dI_k(t)}{dt}\) describes the dynamic amount of infected jackals in an activity center over time for the \(k_{th}\) activity center. It is affected by the following five terms. First, exposed individuals transform into infected individuals at a rate \(\phi\). Second and third, the dispersal between activity centers \(i\) and \(j\) occurs at a rate \(m_{ij}^i\) and with respect to the relative difference in available food. Fourth, infected individuals die at a rate \(\gamma\) due to the virus and at a rate \(\nu_k^i\) from other background reasons.

\begin{equation} \begin{array}{l} \frac{dI_k(t)}{dt} = \underbrace{\phi E_k(t)}_{\text{progression from exposed to infected state}} + \underbrace{\sum_{j=1, j \neq k}^n \big ( m_{jk}^i \frac{I_j(t) \big ( F_k(t) - F_j(t) \big )}{J_k(t) F_k(t)} \big )}_{\text{immigration driven by food gradient}} \ \\\\ - \underbrace{\sum_{j=1, j \neq k}^n \big ( m{kj}^i \frac{I_k(t) \big ( F_k(t) - F_j(t) \big )}{J_j(t) F_k(t)} \big )}_{\text{emigration driven by food gradient}} - \underbrace{\gamma I_k(t)}_{\text{death due to virus}} - \underbrace{\nu_k^i I_k(t)}_{\text{natural death}}. \end{array} \label{eq:i} \end{equation}

Finally, in addition to the dynamics of jackals and rabies, Eq. (\ref{eq:food}), \(\frac{dF_k(t)}{dt}\) describes the dynamics of food availability in each activity center. Food availability is affected by the balance between production and consumption: First, the activity center generates new food with a rate \(a_k\) and up to some amount \(\xi_k\). Second, the local jackal population consumes food at a rate corresponding to the epidemiological classes of the group \(c^s, c^e, \) and \(c^i\) for the susceptible, exposed, and infected, respectively. 

\begin{equation} \begin{array}{l} \frac{dF_k(t)}{dt} = \underbrace{a_k(F_k(t))}_{\text{food production up to carrying capacity}} - \underbrace{c_k^s S_k(t)}_{\text{consumption by susceptible jackals}} - \underbrace{c_k^e E_k(t)}_{\text{consumption by exposed jackals}} - \underbrace{c_k^i I_k(t)}_{\text{consumption by infected jackals}}. \end{array} \label{eq:food} \end{equation}
where \(a_k(F_k(t))\) is a real and positive number \(\lambda \in \mathbb{R}^+\) if \(R(t) < \xi\) for some threshold value \(\xi \in \mathbb{R}^+\) and \(0\) otherwise.  

\subsection{Epizootic intervention policies}
The main motivation for the development of the model describing the epidemiological dynamics above is to investigate the efficiency of different intervention measures. Vaccination and population dilution strategies for jackals are two main epizootic intervention policies (EIPs) in combating the spread of rabies. The vaccines are allocated over time in the jackal's activity centers and are assumed to be consumed randomly by the population. The population dilution is also allocated over time in the jackal's activity centers such that a given number of individuals from the population are removed from the system, independently of the \textit{S/E/I} status. 

In both cases, the EIPs are defined by the amount of vaccines or individuals to remove from each activity center. To this end, both EIPs can be applied more than once at different points in time. Formally, we define the vaccination EIP as a vector \(\mu \in \mathbb{N}^n\) such that each value indicates the number of vaccines distributed in each activity center. The vaccine distribution can occur in multiple points in time \([\tau_1^v, \tau_2^v, \dots, \tau_m^v]\) which define \([\mu_{\tau_1^v}, \mu_{\tau_2^v}, \dots, \mu_{\tau_m^v}]\). In addition, vaccines have a reduced efficacy over time as the vaccine's potency is degraded due to environmental factors, and other (non-target) species might also consume them (reducing overall availability of vaccines for jackals). Similarly, the immunity for rabies inferred by the vaccine also degrades with time to the point that the jackal becomes again susceptible to rabies infection (Eq. (\ref{eq:v_population})). The population dilution EIP is formally defined by \(\psi \in \mathbb{N}^n\) such that each value indicates the number of individuals needed to be removed in each activity center (i.e., the number of shot individuals). The dilution distribution can occur in multiple points in time \([\tau_1^d, \tau_2^d, \dots, \tau_m^d]\) which define \([\psi_{\tau_1^d}, \psi_{\tau_2^d}, \dots, \psi_{\tau_m^d}]\). 

Formally, the amount of vaccination in an activity center, \(k\), is presented in Eq. (\ref{eq:vac_amount}), where \(\frac{dC_k(t)}{dt}\) describes the dynamic amount of available vaccination for jackals in the activity center, over time for the \(k_{th}\) activity center. It is affected by the following three terms. First, vaccinations are allocated at specific times \(\tau_i^v\) such that \(\delta(a, b)\) is the Dirac delta function that defined as:
\[
\delta(a, b) := \begin{cases}
    0, \text{ if } a \neq 0 \\
    b, \text{ if } a = 0 \\
\end{cases}.
\] Second, the jackal population consumes vaccinations at a rate \(\rho\) which corresponds to their epidemiological state (for instance, one can incorporate a situation that sick jackals in state \(I\) differ in their movement, or behavior towards bait). Finally, the vaccination is reduced over time at a rate \(r(t)\) such that \(r(t): \mathbb{R}^+ \rightarrow \mathbb{R^+}\) accepting the temperature in the same time and returns the vaccination reduction rate.

\begin{equation}
    \begin{array}{l}
         \frac{dC_k(t)}{dt} = - \underbrace{ \sum_{i=1}^{m} \delta(t-\tau_i^v, \mu_{\tau_i^v}) }_{\text{vaccination alactivity center}} - \underbrace{\frac{ \sum_{i \in \{s, e, i\}} \rho_i C_k(t) I_k(t)}{J_k(t)}}_{\text{vaccination consumption}} - \underbrace{r(t) C_k(t)}_{\text{effectiveness reduction}}.
    \end{array}
    \label{eq:vac_amount}
\end{equation}

By consuming vaccines, in a random manner, jackals develop immunity which is reduced over time \cite{rabis_israel}. These dynamics are reflected in Eq. (\ref{eq:vac_amount}), where \(\frac{dV_k(t)}{dt}\) describes the dynamic amount of vaccinated jackals in the activity center, over time for the \(k_{th}\) activity center. It is affected by the following three terms. First, individuals in the population become vaccinated at a rate \(\rho\) which corresponds to their epidemiological state. Second, the immunity from the vaccination is reducing over time at a rate \(\omega\) transforming vaccinated individuals back into susceptible individuals. Finally, vaccinated individuals naturally die at a rate \(\nu\). 

\begin{equation}
    \begin{array}{l}
         \frac{dV_k(t)}{dt} = \frac{ \rho_s C_k(t) S_k(t)}{J_k(t)} - \omega V_k(t) - \nu_k^s V_k(t), \\ \\
    \end{array}
    \label{eq:v_population}
\end{equation}

Hence, for the \(k \in [1, \dots, n]\) activity center, the dynamics take the form:
\begin{equation}
    \begin{array}{l}

    \frac{dS_k(t)}{dt} = F_k(t) S_k(t) \big ( 1 - \frac{J_k(t)}{\kappa_k} \big ) - \beta S_k(t) I_k(t)  + \sum_{j=1, j \neq k}^n \big ( m_{jk}^s \frac{S_j(t) \big ( F_k(t) - S_k(t) F_j(t) \big )}{F_k(t)} \big )  - \sum_{j=1, j \neq k}^n \big ( m_{kj}^s  \frac{S_k(t) \big ( F_k(t) - F_j(t) \big )}{J_j(t) F_k(t)} \big )\\ 
    
      - \nu_k^s S_k(t)  - \underbrace{\frac{\rho_s C_k(t) S_k(t)}{J_k(t)}}_{\text{vaccination}} - \underbrace{\sum_{i=1}^{m} \delta(t-\tau_i^d, \psi_{\tau_i^d}) S_k(t)/J_k(t)}_{\text{population dilution}} + \underbrace{\omega V_k(t)}_{\text{vaccination reduced}} ,  \\ \\

    \frac{dE_k(t)}{dt} = \beta S_k(t) I_k(t) - \phi E_k(t)  + \sum_{j=1, j \neq k}^n \big ( m_{jk}^e \frac{E_j(t) \big ( F_k(t) - F_j(t) \big )}{J_k(t) F_k(t)} \big ) \\   
    - \sum_{j=1, j \neq k}^n \big ( m_{kj}^e \frac{ E_k(t) \big ( F_k(t) - F_j(t) \big )}{J_j(t) F_k(t)} \big )  - \nu_k^e E_k(t) - \underbrace{\frac{\rho_e C_k(t) E_k(t)}{J_k(t)}}_{\text{vaccination}} - \underbrace{\sum_{i=1}^{m} \delta(t-\tau_i^d, \psi_{\tau_i^d}) E_k(t)/J_k(t)}_{\text{population dilution}}, \\ \\

    \frac{dI_k(t)}{dt} = \phi E_k(t)  + \sum_{j=1, j \neq k}^n \big ( m_{jk}^i  \frac{I_j(t) \big ( F_k(t) - F_j(t) \big )}{J_k(t) F_k(t)} \big ) - \sum_{j=1, j \neq k}^n \big ( m_{kj}^i \frac{I_k(t) \big ( F_k(t) - F_j(t) \big )}{J_j(t) F_k(t)} \big ) \\ 
    
    - \gamma I_k(t) - \nu_k^i I_k(t) - \underbrace{\frac{\rho_i C_k(t) I_k(t)}{J_k(t)}}_{\text{vaccination}}  - \underbrace{\sum_{i=1}^{m} \delta(t-\tau_i^d, \psi_{\tau_i^d}) I_k(t)/J_k(t)}_{\text{population dilution}},  \\ \\

    \frac{dV_k(t)}{dt} = \frac{ \rho_s C_k(t) S_k(t)}{J_k(t)} + \sum_{j=1, j \neq k}^n \big ( m_{jk}^i  \frac{V_j(t) \big ( F_k(t) - F_j(t) \big )}{J_k(t) F_k(t)} \big ) - \sum_{j=1, j \neq k}^n \big ( m_{kj}^i \frac{V_k(t) \big ( F_k(t) - F_j(t) \big )}{J_j(t) F_k(t)} \big ) - \omega V_k(t) - \nu_k^s V_k(t), \\ \\

    \frac{dF_k(t)}{dt} = a_k(F_k(t)) - c_k^s \big ( S_k(t) + V_k(t) \big ) - c_k^e E_k(t) - c_k^i I_k(t), \\ \\

    \frac{dC_k(t)}{dt} = -  \sum_{i=1}^{m} \delta(t-\tau_i^v, \mu_{\tau_i^v}) - \frac{ \sum_{i \in \{s, e, i\}} \rho_i C_k(t) I_k(t)}{J_k(t)} - r(t) C_k(t),
    \end{array}
    \label{eq:pips}
\end{equation}
where \(J_k(t) := S_k(t) + E_k(t) + I_k(t) + V_k(t)\).

Fig. \ref{fig:scheme} shows a schematic view of the temporal interaction in the model, including the vaccination and population dilution EIPs.

\begin{figure}
    \centering
    \includegraphics[width=0.99\textwidth]{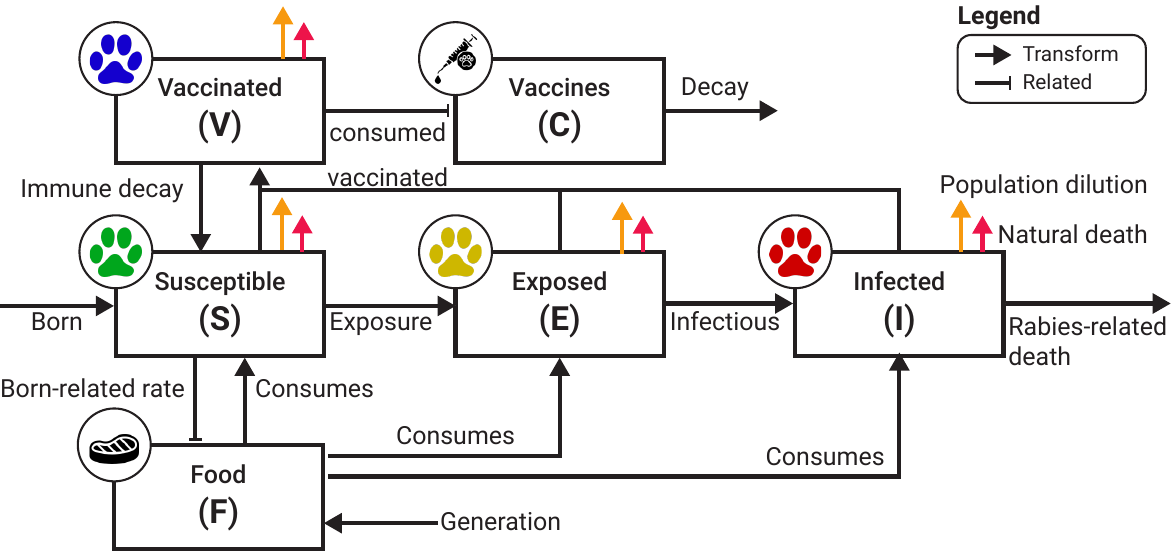}
    \caption{A schematic view of the temporal interaction in the model, including the vaccination and population dilution EIPs.}
    \label{fig:scheme}
\end{figure}

Table \ref{table:params} summarizes the model parameters, grouped into dynamics-related and EIP-related.

\begin{table}[!ht]
\centering
\begin{tabular}{ccp{0.4\textwidth}cc}
\hline \hline
\textbf{Parameter} & \textbf{Category} & \textbf{Description} & \textbf{Value rage} & \textbf{Source} \\ \hline \hline
\(\kappa\) & Dynamics & The carrying capacity of a activity center [1] & 20 - 150 & \cite{rabis_israel} \\ 
\(\beta\) & Dynamics &  The average infection rate of the rabies virus [1] & 0.15 & \cite{DiQuinzio2008} \\ 
\(m_{ij}^x\) & Dynamics &  The average movement rate between activity centers \(i\) and \(j\) in hour [\(t^{-1}\)]& 0.017 & Fitting \\ 
\(\nu^x\) & Dynamics &  The average natural dying rate in hours [\(t^{-1}\)] & \(1.4 \cdot 10^{-5}\) & \cite{Castello2018} \\ 
\(\phi\) & Dynamics &  The average rate of transforming from exposed to an infected individual in hours [\(t^{-1}\)] & \(5.9 \cdot 10^{-3}\) & \cite{exposed_duration} \\ 
\(\gamma\) & Dynamics &  The average dying rate due to the rabies virus in hours [\(t^{-1}\)] & \(9.9 \cdot 10^{-4}\) & \cite{rabis_israel} \\ 
\(c^x\) & Dynamics &  The average food consumption rate in kilograms in hour [\(m/t\)] & \(7.9 \cdot 10^{-3}\) & \cite{Lange2021} \\ 
\(a(\cdot)\) & Dynamics & A function that indicate the food regeneration [\(t\)] & Linear & Assumed \\  
\(T\) & Dynamics & The simulation duration in hours [\(t\)] & 8760 & Assumed \\  
\(\Delta t\) & Dynamics & A simulation step in hours [\(t\)] & 1 & Assumed \\  
\(\omega\) & EIP & The rate vaccination effectiveness reduction rate [1] & 28\% & \cite{rabis_israel} \\   
\(r\) & EIP & The average vaccination effectiveness reduction rate in hours [\(t^{-1}\)]& 120-720 & \cite{vaccine_duration} \\ 
\(\mu\) & EIP & The number of vaccinations distributed in each of the activity centers [\(1\)] & 1-100 & Assumed \\ 
\(\psi\) & EIP & The number of individuals diluted in each of the activity centers [\(1\)] & 1-10 & Assumed \\  
\(\rho^x\) & EIP & The average consumption rate of the vaccination [1] & 10\%-50\% & Assumed \\
\(-\) & EIP & Default vaccines per individual [\(1\)] & 1 & Assumed \\   
\(-\) & EIP & Default dilution portion [\(1\)] & 0.05 & Assumed \\   \hline \hline
\end{tabular}
\caption{The model's parameters with their description. Parameters with the upper index \(x \; (\in \{s, e, i\})\) indicate they are different for the three epidemiological classes (\(S, E, I\)). The value ranges are estimated for the experimental setup (see Section \ref{sec:experiments}).}
\label{table:params}
\end{table}

\subsection{Agent based simulation}
Inspired by previous work \cite{teddy_sir_review}, we opted to implement the model using agent-based simulation (ABS). This approach introduces a higher level of realism by endowing each agent in the population with unique attributes, mirroring the dynamic heterogeneity observed in nature. ABS also mitigates the computational burden of infection computations by employing spatial approximations for interactions between individuals, facilitating the exploration of parameter values and their effects on emerging population-level patterns. 

For the proposed model, each agent in the population (\(a \in J\)) is represented as a tuple $a := (x, y, \eta)$, where $x$ and $y$ are spatial coordinates, and $\eta$ signifies the epidemiological state. Initially, at $t=1$, the overall jackal population is generated based on predefined initial conditions and distributed across a continuous two-dimensional map with dimensions $w$ and $h$. Subsequently, at each time step $t$, each individual is associated with an activity center based on the Euclidean distance (the \(L_2\) metric) it has from the center of the activity center such that the closest activity center is chosen. Interaction between individuals occurs between consecutive time steps based on the activity centers they are associated with, assuming well-mixture \cite{first_sir} inside each activity center. Effectively, this step implies that jackals exist in their activity center, and their movement is condensed to the movement between them. Similarly, it also includes interactions with the EIPs - vaccination and population dilution. Epidemiological dynamics are then applied locally, as represented by Eq. (\ref{eq:pips}). Furthermore, the spontaneous epidemiological processes - exposure to infectious and death due to the virus for infected individuals are computed using time step-associated rates rather than being solely determined by population-level dynamics, aligning with common ABS practices \cite{teddy_pandemic_management}. 

\subsection{Epizootic intervention policy optimization procedure}
\label{sec:opt}
Given the proposed model and its EIPs formalizations, we denote by \(\Upsilon\) the EIP configuration (e.g. the spatio-temporal distribution of vaccines and dilution) used for either of the EIPs, independently of each other. Policymakers (e.g. INPA or managers) are interested in implementing an EIP, \(\Upsilon\), in order to minimize some epidemiological metric \(d\) of the rabies outbreak. For instance, the total number of infected individuals \cite{first_teddy_paper} or the average reproduction number (\(E[R_t]\)) \cite{metric_paper_3}, or any combination thereof (below we specify the specific epizootic indices we have used for optimizing EIPs in our implementation of the model). Formally, the optimization objective (i.e. finding the \(\Upsilon\)) value that will minimize \(d\)) can be expressed as follows \cite{rosenfeld}: 
\begin{equation}
    \min_{\Upsilon} d(\{\mathbb{P}_i\}_{t=1}^{T}). 
    \label{eq:objective}
\end{equation}
In order to solve this optimization task, we first assume individuals are non-strategic, namely, they do not plan their pre-defined path based on the activity center of vaccination or population dilution. We note that because jackals movement decisions between adjacent activity centers are influenced by available resources (\textit{per-capita}), which in turn may depend on EIPs (e.g. dilution), an emergent property of the model is that the jackals realized movements might be indirectly affected by local conditions (e.g. an enhanced movement into diluted location).   

Given a known and fixed number of vaccines (or individual for removal) at the policymaker's disposal (\(b\)), \(\Upsilon\) details where the vaccines will be deployed at each round, namely \(\Upsilon_i \subset V\) such that \(|\Upsilon_i|=k\). Namely, this implies that vaccines are optimally distributed among the different centers. 

A suitable allocation of EIPs can be achieved either optimally or heuristically (both with respect to the distribution of a given EIP effort among activity centers, as well as the relative allocation between them). The simplest heuristic is, obviously, a random allocation of EIPs in the environment. However, such an allocation may be far from optimal. In order to examine the importance of using a strategic allocation policy, in the following evaluation, we use the random allocation mechanism as a null and contrast it with the optimal allocation (in cases where deriving the optimal allocation can be simply performed in reasonable time) or a greedy approximation thereof (otherwise). Specifically, for cases where the number of possible allocations (\(b \choose |V|\)) is less than \(1000\), a simple brute force search was performed. Otherwise, a standard greedy heuristic is used instead where a single EIP allocation is greedily allocated in each iteration, resulting in \(b \cdot |V|\) possible configurations. The exact implementation of these methods is taken from \cite{alexi_games}. Since the greedy approximation of the optimal allocation is, of itself, sub-optimal, using it provides us with a lower bound on the possible competitiveness of the random allocation compared to the optimal one. 

\subsection{Spatio-temporal data fitting procedure}
In order to use the model, one is required to obtain an instance of the model by setting values to the model's parameters, initial condition, and EIP policy. We obtain an instance of the model 
based on the collected data (see Section \ref{sec:data_acquisition}). Formally, given geographical territory, the activity centers of the jackals over time can be determined. This represents a norm-based spatial representation while the proposed model is defined on a graph-based spatial representation. As such, we adopted the method proposed by \cite{norm_to_graph_based} to transfer between the two. Simply put, the fitting procedure generates \textit{in silico} information that agrees with the movement dynamics based on the collected data while also completing the epidemiological dynamics, assuming no EIPs, following Eq. (\ref{eq:pips}). Using this information, the fitting procedure solves an optimization task in which it aims to minimize the number of nodes for a given output graph while also minimizing the population's epidemiological state and spatial distribution differences. To do so, the method can use a variety of heuristic optimization algorithms. In \cite{norm_to_graph_based}, a combination of Time series X-means \cite{tskm_intro} together with multi-agent classification with AutoEncoder \cite{autoencoder,teddy_cancer} are shown to provide the best results, on average, and therefore adopted to this case. Eventually, these diverse algorithms provide the template jackal population for testing the rabies outbreak and optimal EIPs configuration upon. Importantly, as the model's nodes reflect the geographical activity centers from the empirical data and therefore a jackal can not disperse between activity centers that do not share an edge in the graph. Formally, for each activity center (node) \(i \in n\), \(m_{ij} = 0\) if and only if the \(j\) activity center does not share a geographical border between them. 

\subsection{Epizootic spread metrics}
In order to evaluate the epidemic spread, one is required to define an epidemiological metric of interest (i.e. choose the specific \(d\)). In this study, we focused on three popular epidemiological metrics: the average reproduction number (ARN) (\(E[R_t]\)), the maximum number of infections (MI), and the portion of dead individuals in the population due to the epidemic (PDI) \cite{metric_paper_1,first_teddy_paper,metric_paper_2,metric_paper_4}. 

To be exact, \(R_t\) measures the number of secondarily infected individuals given the epidemic state at a given time \(t\) \cite{metric_paper_3}. Hence, the ARN (\(E[R_0]\)) computes how many times, on average, an infected individual infects other individuals. Formally, \(R_t\) can be approximated using the following formula: \(R_t := \big ( I(t) - I(t-1) + R(t) - R(t-1)  \big ) / I(t-1)\). The max infected metric counts the number of individuals infected by the pathogen at a given point in time, divided by the population size (i.e. the proportion of infected individuals). This gives an estimation to the worse \say{wave} of the epizootic spread during the dynamics, and is formally defined at time \(t\) i as follows \(MI(t) := \max_{i \in [t_0, t_f]} I(i)\). Finally, the number of dead individuals due to the epidemic provides insight into the total negative effect that the epizootic has on the population. \(PDI\) at time \(t\) is defined to be the total number of individuals dead divided by the total number of individuals, such that \(t_0\) and \(t_f\) are the beginning and end times of the dynamics, respectively.

\section{Theoretical analysis}
\label{sec:theory}
In this section, we theoretically analyze the proposed model. First, we prove that the model always has a unique solution. Then, we show that population dilution can cause a negative effect on the average epizootic spread while vaccination does not. 

\subsection{Solution existence and uniqueness}
We first show that the proposed model has a solution and it is unique. To this end, we utilize the Picard–Lindelöf theorem \cite{appendix} which states that if \(D \subset \mathbb{R} \times \mathbb{R}^n\) is a closed rectangle with \((t_0, y_0) \in D\) and \(f: D \rightarrow \mathbb{R}^n\) is a function that is continuous in \(t\) and Lipschitz continuous\footnote{A function $f: \mathbb{R}^n \rightarrow \mathbb{R}^m$ is said to be Lipschitz continuous on a set $D \subseteq \mathbb{R}^n$ if there exists a constant $L \geq 0$ such that for all $x, y \in D$, the following inequality holds:
\(\|f(x) - f(y)\| \leq L \|x - y\|, \) where $\|\cdot\|$ denotes the Euclidean norm, and $L$ is referred to as the Lipschitz constant.} in \(y\); then there exists some \(\epsilon > 0\) such that the initial value problem:
\begin{equation}
    y'(t) = f(t, y(t)), y(t_0) = y_0,
\end{equation}
has a unique solution \(y(t)\) on the interval \([t_0 - \epsilon, t_0 + \epsilon]\). Hence, for the proposed model \(y(t) := (S(t), E(t), I(t), V(t), R(t), C(t))\). As such, we use the Picard–Lindelöf theorem by showing that Eq. (\ref{eq:pips}) is continuous in \(t\) and Lipschitz continuous in \(y\). Consequently, let us consider a finite duration in time \([0, T < \infty]\). Following that, the unknown solution, \(y\), has terms of a linear form and of the form \(y_i y_j\). Thus, the function \(f\) such that \(dy(t)/dt = f(t, y(t))\) is continuously differentiable \((C^1)\) which implies that it also locally satisfies Lipschitz condition and continuous in \(t\) \cite{sveta_appendix}. Therefore, we apply the Cauchy–Lipschitz theorem  \cite{final_appendix} which infers that the existence and uniqueness of the solution to
Eq. (\ref{eq:pips}), on any finite interval \([0, T]\). Therefore, a solution exists.

\subsection{Epizootic intervention policies effect}
\label{sec:proof}
The EIPs are designed to reduce the overall spread of an epizootic, it is of interest to explore if the EIP can actually cause more harm than good. To this end, we investigate each of the EIPs (vaccination and dilution) separately, and then, in the experiment section below also their integrative effects in a realized scenario. 

Starting with vaccination, we focus on the average basic reproduction value \((E[R_0])\) as a candidate for the epizootic spread. Let us consider the next-generation matrix \cite{ngm} of the proposed model which infers an average basic reproduction number \(E[R_0] = \frac{\beta - \rho_s}{\phi + \gamma}\). As such, for \(\rho_s > 0\) the vaccination EIP is necessarily reducing the average basic reproduction number and therefore can not cause more harm than good across the entire parameter range. 

Considering the population dilution EIP, we show that - at least theoretically - a negative impact is plausible. Towards this end, let us first consider a simplistic scenario where there are two activity centers with identical population sizes an identical amount of food, and no vaccination. Moreover, the first activity center is full of susceptible individuals only while the latter is full of only infected individuals. We also assume the infection rate \(\beta = 1\) and the mortality rates are \(\nu_s = \nu_i = \gamma = 0\). In such a configuration, without population dilution, or any movement between the two activity centers, the average basic reproduction rate is equal to zero as there are no newly infected individuals (no exposed in the former, and all already exposed in the latter). On the other hand, it is enough to assume a single-time population dilution of a single individual from the former activity center (with susceptible individuals), to initiate a difference in available resources between the two centers. Therefore, at least a single infected individual will disperse from the latter center to the activity center with the susceptible individuals and will infect others. This minimal immigration will bring an infected individual into the susceptible population, resulting in a positive average basic reproduction number. Thus, this scenario demonstrates that population dilution can have a negative impact on the epizootic's spread in certain conditions. To test the relevance of this statement to real-life problems we turn to performing \textit{in silico} experimental setups. 

\section{Experiments}
\label{sec:experiments}
In this section, we describe the experiments conducted using the proposed model and its simulation. Initially, we outline the data acquisition process to obtain realistic spatio-temporal data for the model. Afterward, we formally describe the setup process of our \textit{in silico} experiments. Finally, we present the obtained results.

\subsection{Data acquisition and experimental setup}
\label{sec:data_acquisition}
The golden jackal is a medium-sized omnivorous meso-carnivore canid, that currently expands its distribution range and abundance throughout Eurasia while exhibiting growing densities and in Israel\cite{rabis_israel,doCoutoReis2015}. Being opportunistic in nature, the jackal is a facultative synanthropic species adapted to utilize anthropogenic food resources and thrive in proximity to human settlements or agricultural facilities that create a surplus of resources \cite{Rotem2011}. 

In order to make a realistic configuration of the proposed model and simulation, we collected data for the case of Israel, based on multiple recent reports of rabies in the region \cite{rabis_israel} and the region climate transition from Mediterranean to arid dry-steppe climates. Specifically, the valleys Harod and HaMaayanot (that constitute the study area) are located in the North-East of Israel, connecting between the border with Jordan and the center of Israel. The Harod Valley lies between Mount Gilboa in the south and the Issachar Plateau in the north. HaMaayanot Valley is connected to the Harod Valley on the west and borders in the east with the Jordan River. Both valleys (Harod and HaMaayanot) contain several small rural settlements and agriculture, such as fields, groves, and fisheries. 

Between 2021 and 2024 we trapped 55 jackals within the study area using coil springs traps (Soft catch \#3, Oneida Victor Inc., Euclid, OH, USA). Upon capture, they were carefully transported in a cage to a safe shelter where they were anesthetized by an INPA veterinarian. The jackals' attributes were measured (weight and neck circumference) before being fitted with an ATLAS (Advanced Tracking and Localization of Animals in real-life Systems) collar around their necks. Each jackal was marked by the collar and by a pair of ear tags (Leader large Aussie tags; Bayer AG, Leverkusen, Germany) colored uniquely to correspond to the capture site. 
 
We have used an ATLAS wildlife tracking system to collect the spaio-temporal data of the Jackals. The ATLAS system consists of ground stations with tower-mounted antennas and central data servers \cite{Beardsworth2022}. Tags transmit unique ID signals at 1 Hz, which are received by the stations. If three or more stations detect a signal, the system uses a reverse-GPS method to estimate the tag’s location based on the differential arrival times of the signals \cite{Beardsworth2022, arnon_wildlife_tracking_2023}. While coverage is limited to the study area (unlike global GPS), this method is ideal for tracking local jackals \cite{jw1,jw2,orr_r_4}. It also makes tags inexpensive, lightweight, and battery-efficient, offering good data collection in terms of duration and spatial accuracy \cite{Beardsworth2022}. Notably, lacking empirical data, for simplicity, we assume that food consumption, as well as movement dynamics, are identical between all epidemiological states. 

\subsection{Setup}
\label{sec:setup}
To use the proposed simulation, one is required to set all the model parameters. We adopted the parameter values from Table \ref{table:params}, uniformly randomly sampling the values with ranges to each instance of the simulation. In addition, for the temperature data over time, serving as the base for the vaccine reduction coefficient (\(r(t)\)), we adopted the popular openweathermap API (Application Programming Interface) \cite{weather_api_2,weather_api_1}. Namely, to obtain a representing temperature for each day, we averaged the temperature of each day of the year based on the last decade. 

Based on this setup, we conducted three main experiments for the proposed model. As the number of parameters and their values can range widely from one instance of the model to another, we computed the model for \(n = 100\) simulation repetition, each randomly sampling the values of the parameters in the model using a uniform distribution across feasible ranges as summarized in Table \ref{table:params}. A visualization of the jackals' empirically calculated (from tracking data) activity centers is provided in the Appendix. Hinged on these spatio-temporal settings, we first investigate seven different EIP configurations: no EIPs,  random vaccination (without dilution), random dilution (without vaccination), random both, optimal vaccination (alone), optional dilution (alone), and optimal both. The no EIP is the baseline case where no EIP is utilized. The three random EIP configurations utilize the default EIP values (see Table \ref{table:params}) and use them in a completely random manner. Finally, the three optimal EIP configurations utilize the default EIP values following an optimal policy obtained using the optimization procedure (see Section \ref{sec:opt}). 

\subsection{Results}
\label{sec:results}
\begin{figure}
    \centering
    \includegraphics[width=0.99\textwidth]{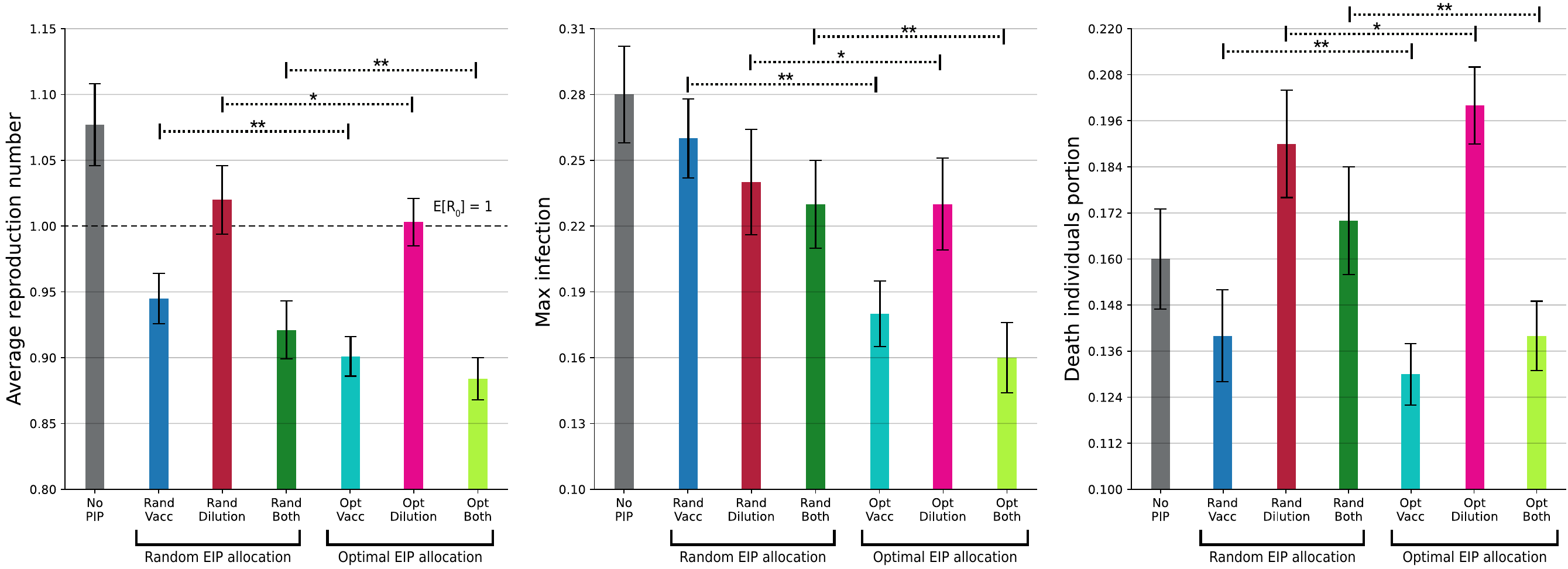}
    \caption{A comparison between different EIP strategies in terms of the three epizootic spread metrics. The results are shown as the mean \(\pm\) standard deviation of \(n=100\) repetitions. }
    \label{fig:pip_differances}
\end{figure}

Fig. \ref{fig:pip_differances} presents a comparison between seven EIP configurations in terms of the three indices of the epidemic spread metrics - average reproduction number (ARN (\(E[R_0]\))), max infected (MI), and the portion of dead individuals (PDI). The results are shown as the mean \(\pm\) standard deviation of \(n=100\) simulation repetitions. Broadly speaking deploying EIPs reduces both ARN and MI metrics, while, this is not the case for PDI. Comparing the no EIP configuration to the EIP configuration for all three metrics using Kruskal-Wallis H Tests, reveals that the no-EIP is statistically significantly worse with \(p < 0.01\) and \(p < 0.05\) for the ARN and MI, respectively, while not for the PDI metric, respectively. Similarly, comparing each pair of random and optimal EIP configurations with Mann‐Whitney U tests \cite{mcknight2010mann} reveals that for all three cases (only dilution, only vaccination, and both) the optimal EIP configuration is statically significantly better in preventing epizootic with all three metrics with \(p < 0.01, p < 0.05,\) and \(p < 0.05\), respectively. Overall, we find that for ARN and MI including any of the two EIPs (vaccination, dilution), or their combination reduces outbreak magnitude, and that optimal EIPs tend to be better than their random application. For PDI, in contrast, we find that while vaccination is always helpful, EIPs including dilution (either random, optimal, or together with vaccination) result in a significantly higher effect of the disease on the population. The reason for these different dynamics of EIP usage in terms of the ARN and MI compared to PDI, is that the PDI counts all deaths. Thus, for PDI the dilution of susceptible jackals is also counted, contributing to overall higher death proportion, regardless the net effect of the epidemic. Importantly, the figure shows an agreement with the general theoretical solution - the vaccination EIP (either random or optimal configuration) is able to reduce all three epizootic spread metrics, while dilution does not. In fact, in a broad range of conditions, dilution is actually having the opposite result (i.e., contributing to the epizootic spread), in agreement with the theoretical solution shown in section \ref{sec:theory}. This antagonistic effect of dilution highlights the necessity to further investigate these dynamics, and identifying relevant conditions of this effect.

\begin{figure}
    \centering
    \includegraphics[width=0.99\textwidth]{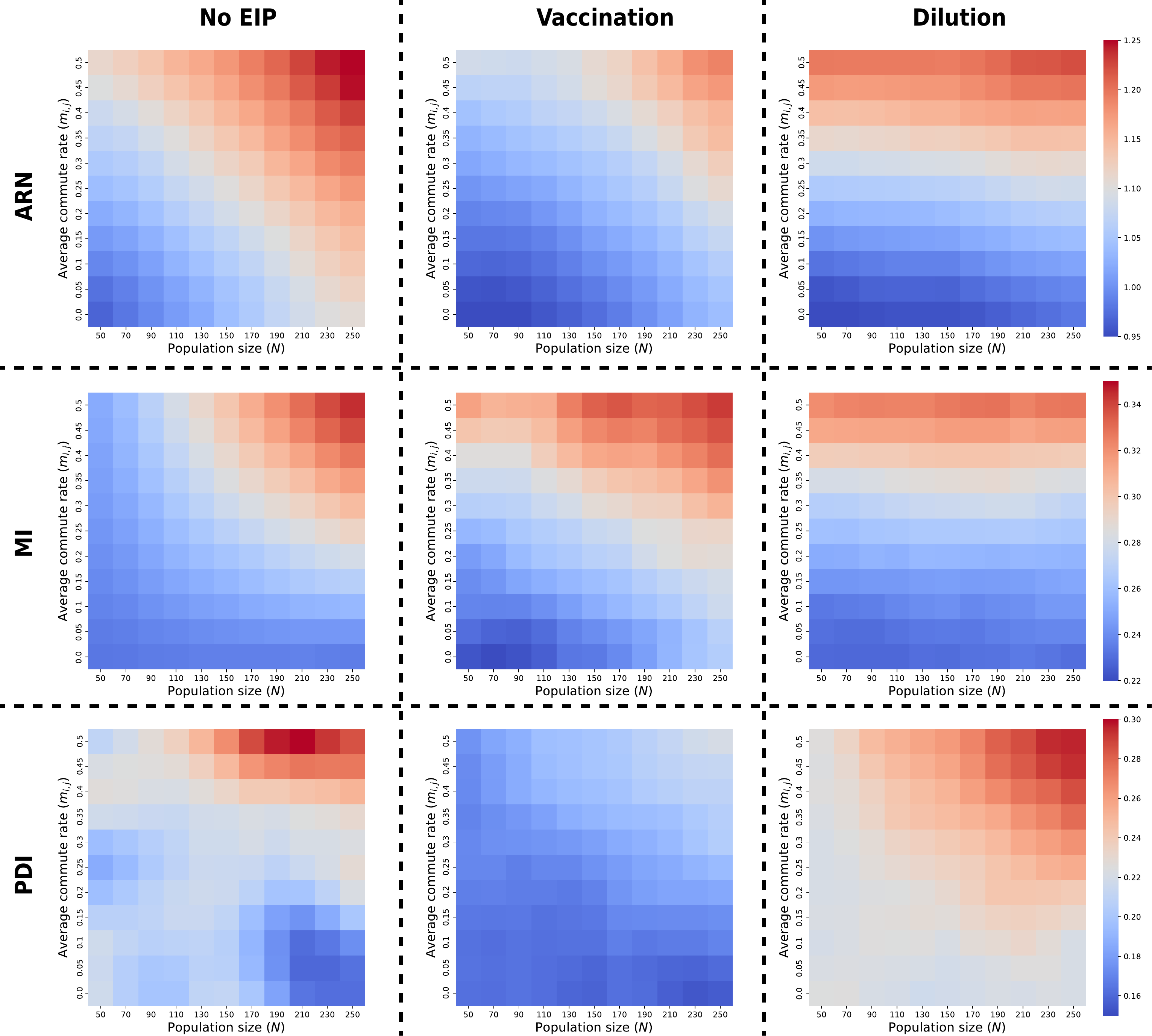}
\caption{The influence of population size (the x-axis) and immigration rate (y-axis, i.e., dispersal between activity centers) on the epizootic spread under an optimal EIP configuration. The results are shown as the mean of \(n=100\) repetitions for each configuration (no EIP, vaccination only, and dilution only) with warmer colors indicating higher values of epizootic indices for ARN, MI, and PDI in the three respective rows.}
\label{fig:heatmaps}
\end{figure}

To further investigate how the population size and individuals' local immigration (between adjacent activity centers) interact to affect the disease spread,  we used the optimal EIP strategy configuration. Fig. \ref{fig:heatmaps} summarizes the results of this analysis, showing the values as the mean outcome of \(n = 100\) simulation repetitions on the three rabies epizootic indices (ARN, MI, and PDI). Specifically, with an increase in both population size (x-axis) and in the average immigration rate (y-axis), the ARN grows monotonically for the No-EIPs scenario. This increase is due to higher \(\beta\) values because of improved mobility, and due to higher densities in the two axis, respectively. Without EIPs the epizootic will spread (\(R_0>1\)) for most of the considered predictors' range. Including optimal vaccination results in a similar trend (ARN increases with both axes), but with overall much lower values (colder colors overall), implying the epizootic will decay in low densities and immigration rates, and will only mildly spread in the very high values. Lastly, with optimal dilution, ARN shows different behavior, with no influence of population size (because the same fraction will be diluted) but a high sensitivity to immigration rate. Overall ARN values are much higher in this scenario, indicating effective epizootic spread despite this EIP effort. This result reflects the limited efficiency of the dilution as EIPs, even if performed optimally, as well as its sensitivity to local host movements, and the epizootic spread due to the jackals' tendency to immigrate to less occupied activity centers.    

Similar patterns have emerged for the MI metric, as presented in the second row of Fig. \ref{fig:heatmaps}, these patterns entail that no EIP results in high mortality with increasing values with both predictors and dilution being affected only by migration rates and not by population density. Interestingly, in the case of vaccination only, despite the general trend of a combined effect of both predictors, high immigration values result in very high mortality compared both to the No-EIP and to the dilution-only scenarios. The reason for this result is that while vaccinated individuals slow the epizootic spread, the accumulating number of infected individuals over time is larger, similar to other EIPs that delay infection dynamics \cite{teddy_economy}. 

Unlike these two metrics (ARN and MI), for the PDI metric (lower row), the results indicate somewhat different epizootic dynamics, with unstable trends indicated in less clear gradients across the heatmaps. Specifically, in contrast to the additive effect above, the two predictors strongly interact in the No-EIPs scenario: the effect of immigration rate is quite unimportant for small populations (with low density values), but very pronounced in dense populations, ranging from very low to very high PDIs as mobility grows in a dense population. The results show the strong efficiency of vaccination in reducing relative mortality rates, effectively keeping them low (below the median value) across the entire parameter range, with only minor enhancement by increasing density and movement. Finally, the dilution EIP results in higher PDI across the board compared to the vaccination EIP, with a strong increase in relative mortality as the density and movement grow. These results can be explained by the removal of individuals by dilution as well as by the fact that dilution strongly facilitates movement of individuals between activity centers, which results in a rise in epizootic spread leading to higher overall mortality in the long run. 

Taken jointly, comparing optimal vaccination and dilution, highlights the superiority of the former across a broad range of ecological conditions, with lower ARN values and relative rabies-induced mortality. Further, dilution also showed a stronger sensitivity to increasing movement rates (because individuals immigrate according to resource availability, therefore they readily arrive to fill the vacancy created by dilution). Dilution also showed a detrimental effect of proportional mortality (even if the maximal disease mortality is not higher the population is smaller due to the additional removal of otherwise healthy individuals).   

\begin{figure}[!ht]
    \centering
    \includegraphics[width=0.99\textwidth]{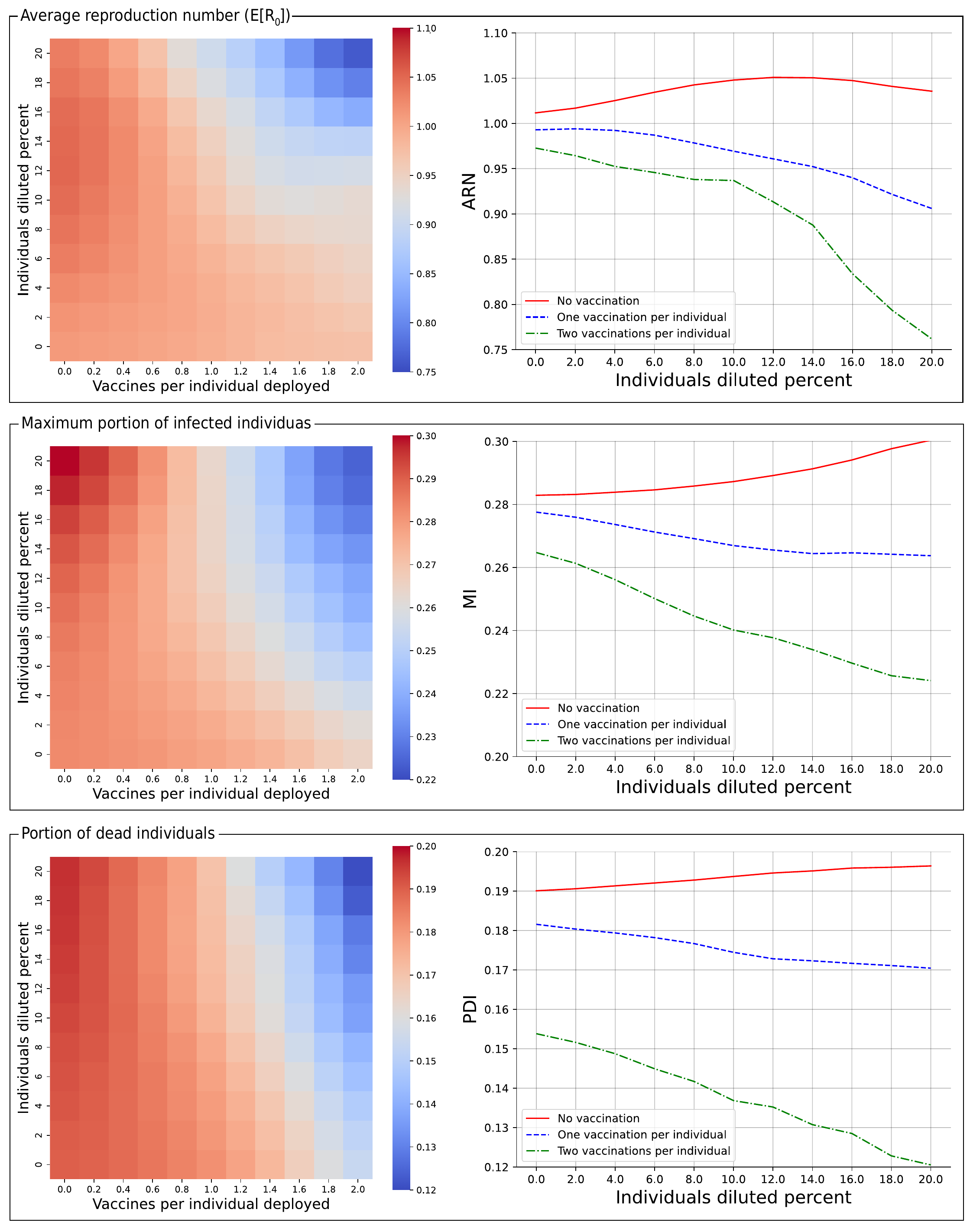}
\caption{The influence of different EIP combinations under the optimal configuration. The results are shown as the mean of \(n=100\) repetitions for each configuration.}
\label{fig:combined_pip}
\end{figure}

In a similar manner, Fig. \ref{fig:combined_pip} outlines the influence of different EIP combinations under the optimal configuration on the three epizootic indices, showing the values as the mean outcome of \(n = 100\) simulation repetitions. Focusing on the ARN, we find that with no or partial vaccination, increasing dilution (until intermediate levels of around 10\% diluted) has a counter-productive effect of enhancing ARN. This response is absent with high vaccination rates where increasing dilution has a monotonic negative effect on ARN (i.e. the desired effect of reduction in the epizootic across all dilution values; the top-left cell in the heatmap). This interaction is a non-trivial prediction provided by our models, showing that the efficiency of a secondary EIP (dilution), qualitatively and strongly depends on the coverage by the primary EIP of vaccination. Overall, the larger the amount of vaccines deployed, the lower the ARN is predicted to be, with a non-linear relationship to the amount of dilutions. This dynamic is also reproduced (and even enhanced) for MI and PDI as presented in the second and third rows, respectively. At low vaccination rates increasing dilution leads to higher (rather than lower) mortality, in contrast to the desired effect (and while this effect was limited to intermediate dilution values for the ARN, here it is prevailing up to 20\% dilution ). Only when vaccination rates are high (more than \(0.8\) vaccines per individual), dilution contribute monotonically to the mitigation of the disease. 

We suggest that two factors contribute to this counter-intuitive dynamics, and antagonistic effect of dilution at low vaccination levels. First, as described above, by reducing activity-center-specific populations, the dilution facilitates movement among them, thus contributing to the encounters of susceptible and infected individuals. Second, the removal of vaccinated individuals may have a stronger effect where these are still rare, contributing to encounters that translate to actual transmission. Taken together, these results provide a quantitative evaluation of the potential interference between complementary EIPs, and highlight the dependency of dilution efficiency in vaccination prevalence.      

\section{Discussion}
\label{sec:discussion}
In this study, we propose a novel spatio-temporal epidemiological model for the spread of rabies using data from a population of golden jackals (known to be the main host in the focal system \cite{gdalevich2000rabies,rabis_israel}). The proposed model is based on several extensions of the well-established SIR modeling approach \cite{jones2021spread,sene2020sir,ahmed2023bifurcation} into a spatially-explicit graph-based model and a Susceptible-Exposed-Infected (SEI) temporal framework (reflecting the virus biology and incubation period). We then implement the model as an agent-based simulation approach using the data collected from the study area in Harod Valley (telemetry data of jackals' movements). Other relevant epidemiological parameters are obtained from the biological literature (for instance, the rabies infection rate in jackals). This model fitting on a realistic setup, allows us to investigate \textit{in silico} the \textit{in situ} biological scenarios. In particular, we use our model and the associated simulations to explore the efficiency of epizootic intervention policies (EIPs), focusing on the additive and interactive effects of vaccination and dilution, which are the two most commonly applied EIPs for mitigating rabies outbreaks \cite{Rupprecht2024}. We find that while vaccination has a desired (epizootic mitigation) effect throughout the relevant parameter hyper-range, this is not the case for the dilution EIP. The theoretical argument shows dilution may result in a counter-productive effect (i.e. enhancing disease spread). Indeed, the simulations demonstrate that this general notion is expected for biologically relevant scenarios - namely, dilution applied with insufficient vaccination may result in more pronounced outbreaks. Below, we first overview the basic model findings, then we discuss how ecological conditions affect the effectiveness of the two EIPs and their mutual interference. We end our discussion by highlighting a few shortcomings of the current models and pointing out possible future directions and applications for management. 

\subsection{Model insights and ecological factors affecting rabies epizootic dynamics}
Our model generates some trivially expected patterns, that demonstrate its relevance and biological robustness, despite the broad range of values used for some of the less explored parameter hyperspace (lacking narrow, and accurate system-specific information). First, without any EIPs, the epizootic spreads in the population with a positive ARN (e.g. \(E[R_0] = 1.18\)), and infection-derived mortality is around 28\%  of the population. Second, as one may expect \cite{teddy_pandemic_management}, the model shows (Fig. \ref{fig:pip_differances)} that the two EIPs, when applied separately, reduce epizootic spread across a broad range of conditions. Third, we find that an optimal application of each EIP is more effective than a null hypothesis of random application which fails to account for the specific conditions. For instance, when vaccination is utilized in an optimal manner, the expected reduction in the epizootic spread, in terms of the ARN is 0.36, reducing it well under the \(E[R_0] = 1\) threshold and ensuring a rabies epizootic to end relatively quickly. The random vaccination, however, will have a weak effect on ARN of merely ~0.94. Importantly, the sensitivity analysis, as embellished by Fig. \ref{fig:sensitivity}, shows that the proposed results are relatively stable with a local (semi-)linear change in the ARN with respect to various properties, further supporting the effectiveness of the proposed model and simulation. Fourth, we find that, on average, the combined vaccination and population dilution reduce rabies spread more than each one independently, as shown in Fig. \ref{fig:pip_differances}. Overall, while these specific predictions of our model are realistic but not innovative, they allow us to turn to investigate our questions of interest regarding the effects of ecological conditions (namely jackals' movement rates and population density) on the efficiency of the two EIPs, and their interactive effects, and their possible interference \cite{teddy_pandemic_management}. 

A large body of literature emphasizes the importance of host behavior and ecology for epizootic dynamics \cite{Dougherty2022,Dobson1986}. Dense populations almost invariably facilitate spread by enhancing contact rates (\(\beta\)) and duration among infected and susceptible hosts \cite{Daszak2000}. In India for instance, over-abundant resources (due to the disappearance of natural scavengers) have led to a very high increase in the density of stray dogs, resulting in massive rabies outbreaks, and surplus human mortality \cite{Frank2023,Frank2024}. Locally, frequent rabies outbreaks in northern Israel, and the central role jackals play in them as hosts, have been attributed to the growing density of their population. The plasticity of the golden jackal, being a synanthrope with an ability to utilize anthropogenic waste allows them to competitively exclude foxes from large areas, where they previously dominated reported rabies incidence \cite{rabis_israel} These notions are well supported in our results, where denser jackal populations are enhancing outbreaks (higher ARN and MI) and limited efficiency of the EIPs at higher densities (Figure 5). 

In addition to host density, also host movement patterns are subject to well-developed ecological studies in the context of disease epidemiology \cite{Spiegel2022,Anglister2024}. Broadly speaking, more pronounced movement contributes to faster disease spread in space as well as among hosts. This is also true for our model, where jackals' tendency to immigrate among adjacent activity centers strongly affects outbreak magnitude and the EIPs efficiency. Interestingly, the movement predictor (which is often less addressed compared to the density), had stronger effect on the outbreak indices. The PDI index (proportion of dead overall) demonstrates it clearly if no EIPs are used, showing an interactive effect: with low movement rates, increasing density actually has a negative impact on the ARN (Fig 5 lower left). Despite similar death rates for dense populations (MI), the relative effect on the population is smaller, presumably because the epizootic often \say{dies} within an activity center before being spread by emigrating jackals. In contrast, when movement rates are high, we see a positive increase in PDIs with population density, because the disease readily spreads among centers, resulting in higher death rates and more acute outbreaks. Overall, these results concur with the growing awareness in the field \cite{Ezenwa2016,Dougherty2022}, highlighting the potential of host movement to govern qualitative patterns of outbreak dynamics, beyond a simple quantitative intensification. 

\subsection{Possible interference between epizootic intervention policies }
When further exploring how the optimal EIP configurations perform under different movement patterns and population sizes. The simulations show that there is a significant impact on the spatial distribution of infection, as illustrated by Fig. \ref{fig:heatmaps}. This finding is consistent with empirical studies on pandemic and epizootic spread in both humans and animals \cite{wong2020spreading,moreno2023stocking,yin2021association}. Traditional models suggest that vaccination and reduced density independently slow transmission \cite{wong2020spreading}, and this was also supported by recent empirical reviews in the context of rabies \cite{Rupprecht2024}. The effectiveness of dilution, is more controversial \cite{Morters2013}, and our model reveals that even an optimal application of dilution might have limited efficiency in some conditions. Further, because optimal dilution depends on accurate data on local densities of the various activity centers, as well as data on their movement patterns, achieving this goal is practically unfeasible in field conditions, suggesting that effectively, dilution will be even less efficient in epizootic mitigation. Another challenge that suggests dilution is likely to have limited efficiency as EIP in real-life situation is pointing to the required effort. While we find that the efficacy of optimal dilution was similar across the range of growing population density, it should be noted that diluting a similar proportion in larger populations requires the removal of a growing number of individuals, which is likely to be a practical methodological challenge in a real-life context.  

A particularly novel and important finding from this study is that certain combinations of vaccination and population dilution can be anatogonistic, exacerbating rabies spread under some scenarios. This outcome is supported by both a theoretical (see Section \ref{sec:proof}) and empirical (see Fig. \ref{fig:combined_pip}) analysis of the proposed model and data. The simulations reveal that dilution effort without adequately coordinated effective vaccination efforts, results in higher ARN, MI, and PDI values. This outcome reflects the limitation of the dilution EIP due to the lack of discrimination between individuals in different epidemiological states (potentially removing also vaccinated individuals), and its effect on the population's local densities, which in turn, may alter dispersal dynamics. By removing individuals, the dilution may increase relative resource availability in these areas, encouraging immigration among centers (and the spread of the disease, if some of the dispersing jackals was exposed to rabies). This counterintuitive result suggests that, under biologically relevant ecological conditions, an unbalanced intervention approach could unintentionally sustain or even amplify disease persistence, an outcome policymakers should take into consideration when designing a EIP for each specific scenario. Obviously, specific value ranges for such antagonistic results by dilution will depend on local biology, and additional behavioral parameters not covered by our model (e.g. territoriality that is common in many canids \cite{moorcroft2006mechanistic,potts2012territorial}, age-dependent movement and other behavioral mechanisms). Nevertheless, empirical studies have highlighted how disruption of the social structure may contribute to movements and disease \cite{rozins2018social,viana2023effects,downing2023culling}. 
Accordingly, we suggest that the general antagonistic effect dilution to vaccination efforts is likely to prevail in many systems where both EIPs are lower than optimal due to the practical challenges involved.

\subsection{Study limitations, future directions and concluding remarks}
While our model presents a comprehensive approach to studying the mitigation of rabies epizootic spread in jackal population using population dilation and vaccinations, several limitations of the proposed model and analysis should be discussed. First, like other epidemiological models \cite{cooper,sir_example_2,dang}, the proposed model contains a few biological simplifications, reflecting a compromise between realism and generality. Overly complex model might be limited by the data availability and reliability for many of their parameters, and suffer from lower transferability across biological systems. Model complexity may also prove to be extremely time and resource-consuming to obtain and parametrize \cite{levy2015modeling}. Second, another limitation in the application of our ABSs is the coverage of the hyper-parameter space. Uniformly sampling from the entire range might under-represent more feasible values of the parameters, contributing to variation among specific iterations, and limiting the biological value of the models. Future studies can strive to obtain more accurate values for the various parameters, and improve sensitivity analyses for their respective influence.     
Improving the biological accuracy and robustness of the model in future work may strive to further attune some of the simplifications assumed in the current model. First, while the jackal population is assumed to be homogeneous, age-dependent behavior, as well as consistent intraspesific differences in movement tendencies and behaviors are both prevailing in wild populations \cite{spiegel2017s}, and establish factors affecting disease dynamics \cite{teddy_cows} which can improve model accuracy. 
Second, jackals' movement in our model is reduced to the single axis of immigrating among activity centers. Jackals' social behaviors and structures (e.g. packs and territoriality) are ignored despite their potential influence on infection rates within each activity center. Recent works show that the infection rate is heavily influenced by the potential interactions within different activity centers and the ecological properties (such as resource distribution inside the activity center) \cite{large_models_not_fit,teddy_cows,teddy_ariel}. Third, despite clear evidence supporting behavior alternation by rabies infection, we do not account for the potential impacts of the infection status on other behaviors such as immigration rates or infection rates (throughout increased aggressiveness) \cite{sick_walk_2}. Measuring if infected hosts differ in these behaviors and including this disease-dependent rates, can provide qualitatively different predictions. Assuming that only commuting between activity centers reflects the behavioral aspect of the jackals, while ignoring other behaviors such as hunting \cite{vsprem2024factors}. Second, we assume the infection rate (\(\beta\)) is constant between activity centers as a property of the pathogen \cite{or_sir_example,con_new_2,intro_1}.
Fourth, the model focuses on a single species (i.e., jackals), ignoring cross-species epizootic dynamics \cite{base_paper}, in particular owned-dogs and stray-dogs which are prominent actors in rabies spread in Isreal\footnote{\url{https://www.gov.il/he/pages/rabies-occurrence-years?chapterIndex=12}}.
Finally, we assume individuals are aware of the food consumption rates in all adjacent activity centers. This assumption might be false for even relatively small sub-populations as previous studies show animals perform exploration-exploitation behavior, and that their movement and information status are interconnected  \cite{lazebnik2024exploration,eve_nature_1,eve_nature_4,eve_nature_5,SPIEGEL201690}. Future work should relax one or more of these assumptions to improve the model's realism and obtain more accurate predictions for different EIP configurations. 

Taken jointly, our model underscores the utility of integrating extended SIR models and empirically parameterized simulations to address concurrent challenges related to disease transmission in animal populations \cite{con_new_6}. While collecting data on real-life movement patterns of free-ranging animal hosts may challenge model parametrization, our model, benefiting from a large-scale jackal tracking effort demonstrates the potential contribution of this combined approach, and potential insights into disease outbreak and mitigation. With the rapid improvement in animal tracking technologies \cite{orr_end_1}, accumulating empirical case studies are rapidly closing this gap \cite{Barrile2024,Grabow2024} between movement ecology and epidemiology \cite{Dougherty2022}. By augmenting these models with empirical data, we can improve predictions concerning specific systems and pathogens. One counter-intuitive result our analysis reveals is that population dilution, when performed randomly, can increase the epizootic spread over time, on average, with or without a parallel vaccination effort. Moreover, we show that even for optimal EIP configuration, a population's movement plays a critical role in the EIP effectiveness, and in general higher mobility leads to reduced EIP effectiveness. Given the rapid increase in zoonotic disease prevalence and the management efforts to mitigate these \cite{orr_3}, policymakers should account for these outcomes when dealing with similar scenarios.

\section*{Declarations}
\subsection*{Funding}
OS acknowledges financial support by a grant from the Israeli National Parks Authority (2021); from the Israeli Science Foundation (ISF396/20 and 505/24); and from the Open Space Preservation Fund of the Israel Land Authority (2019, 2023). 

\subsection*{Conflicts of interest/Competing interests}
None.

\subsection*{Data Availability}
The data that have been used in this study is available upon reasonable request from the authors.
 
\subsection*{Contribution statement}
Teddy Lazebnik: Conceptualization, Resources, Software, Formal Analysis, Investigation, Methodology, Visualization, Project administration, Writing - Original Draft, Writing - Review \& Editing. \\  
Yehuda Samuel: Data curation, Formal Analysis, Writing - Review \& Editing. \\
Roi Lapid: Data curation. \\
Roni King: Data curation. \\
Erez Ben Yosef: Data curation. \\
Orr Spiegel: Conceptualization, Data curation, Validation, Supervision, Investigation, Writing - Review \& Editing. \\
 
\bibliography{biblio}

\begin{thebibliography}{100}

\bibitem{orr_1}
S.~J. Salyer, R.~Silver, K.~Simone, and C.~Barton~Behravesh.
\newblock Prioritizing zoonoses for global health capacity building-themes from one health zoonotic disease workshops in 7 countries, 2014-2016.
\newblock {\em Emerging Infectious Diseases}, 23(13):S55--S64, 2017.

\bibitem{orr_2}
M.~E.~J. Woolhouse, C.~Dye, S.~Cleaveland, M.~K. Laurenson, and L.~H. Taylor.
\newblock Diseases of humans and their domestic mammals: pathogen characteristics, host range and the risk of emergence.
\newblock {\em Philosophical Transactions of the Royal Society of London Series B: Biological Sciences}, 356(1411):991--999, 2001.

\bibitem{intro_2}
L.~A. White, J.~D. Forester, and M.~E. Craft.
\newblock Disease outbreak thresholds emerge from interactions between movement behavior, landscape structure, and epidemiology.
\newblock {\em Proceedings of the National Academy of Sciences}, 115(28):7374–7379, 2018.

\bibitem{orr_3}
M.~E. Craft.
\newblock Infectious disease transmission and contact networks in wildlife and livestock.
\newblock {\em Philosophical Transactions of the Royal Society B: Biological Sciences}, 370(1669), 2015.

\bibitem{con_new_5}
J.~M. Hassell, M.~Begon, M.~J. Ward, and E.~M. Fevre.
\newblock Urbanization and disease emergence: Dynamics at the wildlife–livestock–human interface.
\newblock {\em Trends in Ecology \& Evolution}, 32(1):55--67, 2017.

\bibitem{orr_4}
J.~S. Mackenzie, M.~Jeggo, P.~Daszak, and J.~A. Richt.
\newblock {\em One Health: The human-animal-environment interfaces in emerging infectious diseases}.
\newblock Springer, 2013.

\bibitem{intro_1}
E.~R. Dougherty, D.~P. Seidel, C.~J. Carlson, O.~Spiegel, and W.~M. Getz.
\newblock Going through the motions: incorporating movement analyses into disease research.
\newblock {\em Ecology}, 2017.

\bibitem{intro_0}
O.~Spiegel, N.~Anglister, and M.~M. Crafton.
\newblock {\em Movement data provides insight into feedbacks and heterogeneities in host-parasite interactions}, pages 91--110.
\newblock 2022.

\bibitem{con_new_3}
V.~O. Ezenwa, E.~A. Archie, M.~E. Craft, D.~M. Hawley, L.~B. Martin, J.~Moore, and L.~White.
\newblock Host behaviour–parasite feedback: an essential link between animal behaviour and disease ecology.
\newblock {\em Proceedings of the Royal Society B: Biological Sciences}, 283(1828):20153078, 2016.

\bibitem{intro_3}
S.~Eubank, H.~Guclu, V.~S. Anil~Kumar, M.~V. Marathe, A.~Srinivasan, Z.~Toroczkai, and N.~Wang.
\newblock Modelling disease outbreaks in realistic urban social networks.
\newblock {\em Nature}, 429(6988):180–184, 2004.

\bibitem{lv_pandemic_3}
B.~Sahoo and S.~Poria.
\newblock Disease control in a food chain model supplying alternative food.
\newblock {\em Applied Mathematical Modelling}, 37:5653–5663, 2013.

\bibitem{lv_pandemic_4}
Z.~Sabir, T.~Botmart, M.~A.~Z. Raja, and W.~Weera.
\newblock An advanced computing scheme for the numerical investigations of an infection-based fractional-order nonlinear prey-predator system.
\newblock {\em Plos One}, 17(3):e0265064, 2022.

\bibitem{teddy_multi_strain}
T.~Lazebnik and Bunimovich-Mendrazitsky S.
\newblock Generic approach for mathematical model of multi-strain pandemics.
\newblock {\em Plos One}, 17(4):e0260683, 2022.

\bibitem{multi_populations_1}
V.~Ram and L.~P. Schaposnik.
\newblock A modified age-structured sir model for covid-19 type viruses.
\newblock {\em Scientific Reports}, 11:15194, 2021.

\bibitem{multi_populations_4}
I.~E. Marie and K.~Masaomi.
\newblock Effects of metapopulation mobility and climate change in si-sir model for malaria disease.
\newblock page 99–103. Association for Computing Machinery, 2020.

\bibitem{multi_populations_3}
A.~J. Terry.
\newblock Pulse vaccination strategies in a metapopulation sir model.
\newblock {\em Mathematical Biosciences \& Engineering}, 7(2):455--477, 2010.

\bibitem{first_teddy_paper}
T.~Lazebnik and S.~Bunimovich-Mendrazitsky.
\newblock The signature features of covid-19 pandemic in a hybrid mathematical model—implications for optimal work–school lockdown policy.
\newblock {\em Adv. Theory Simul.}, 4(5):e2000298, 2021.

\bibitem{different_approach_from_sir}
B.~Ivorra, M.~R. Ferrandez, M.~Vela-Perez, and A.~M. Ramos.
\newblock Mathematical modeling of the spread of the coronavirus disease 2019 (covid-19) taking into account the undetected infections. the case of china.
\newblock {\em Commun Nonlinear Sci Numer Simulat}, 2020.

\bibitem{different_approach_from_sir_3}
L.~Nesteruk.
\newblock Statistics-based predictions of coronavirus epidemic spreading in mainland china.
\newblock {\em Innov Biosyst Bioeng}, 4:13--18, 2020.

\bibitem{different_approach_from_sir_2}
J.~B. Long and J.~M. Ehrenfeld.
\newblock The role of augmented intelligence (ai) in detecting and preventing the spread of novel coronavirus.
\newblock {\em Journal of Medical Systems}, 44, 2020.

\bibitem{intro_models_1}
L.~Wang, T.~Xu, T.~Stoecker, H.~Stoecker, Y.~Jiang, and K.~Zhou.
\newblock Machine learning spatio-temporal epidemiological model to evaluate germany-county-level covid-19 risk.
\newblock {\em Machine Learning: Science and Technology}, 2(3), 2021.

\bibitem{orr_r_1}
J.~O. Lloyd-Smith, S.~J. Schreiber, P.~E. Kopp, and W.~M. Getz.
\newblock Superspreading and the effect of individual variation on disease emergence.
\newblock {\em Nature}, 438(7066):355--359, 2005.

\bibitem{teddy_pandemic_management}
T.~Lazebnik, S.~Bunimovich-Mendrazitsky, and L.~Shami.
\newblock Pandemic management by a spatio–temporal mathematical model.
\newblock {\em International Journal of Nonlinear Sciences and Numerical Simulation}, 107(4):106176, 2021.

\bibitem{first_sir}
W.~O. Kermack and A.~G. McKendrick.
\newblock A contribution to the mathematical theory of epidemics.
\newblock {\em Proceedings of the Royal Society}, 115:700–721, 1927.

\bibitem{bio_2}
V.~Piccirillo.
\newblock Nonlinear control of infection spread based on a deterministic seir model.
\newblock {\em Chaos, Solitions \& Fractals}, 149:111051, 2021.

\bibitem{spatial_2_example}
A.~Viguerie, G.~Lorenzo, F.~Auricchio, D.~Baroli, T.~J.~R. Hughes, A.~Patton, A.~Reali, T.~E. Yankeelov, and A.~Veneziani.
\newblock Simulating the spread of covid-19 via a spatially-resolved susceptible–exposed–infected–recovered–deceased (seird) model with heterogeneous diffusion.
\newblock {\em Applied Mathematics Letters}, 111:106617, 2021.

\bibitem{multi_strain_3}
O.~Khyar and K.~Allali.
\newblock Global dynamics of a multi-strain seir epidemic model with general incidence rates: application to covid-19 pandemic.
\newblock {\em Nonlinear Dynamics}, 102:489--509, 2020.

\bibitem{pip_5}
M.~R. Taylor, K.~E. Agho, G.~J. Stevens, and B.~Raphael.
\newblock Factors influencing psychological distress during a disease epidemic: Data from australia's first outbreak of equine influenza.
\newblock {\em BMC Public Health}, 8:347, 2008.

\bibitem{pip_2}
M.~I. Meltzer, N.~J. Cox, and K.~Fukuda.
\newblock The economic impact of pandemic influenza in the united states: priorities for intervention.
\newblock {\em Emerging Infectious Diseases}, 5(5):659--671, 1999.

\bibitem{pip_3}
M.~Kabir, M.~S. Afzai, A.~Khan, and H.~Ahmed.
\newblock Covid-19 pandemic and economic cost; impact on forcibly displaced people.
\newblock {\em Travel Medicine and Infectious Disease}, 35:101661, 2020.

\bibitem{pip_4}
P.~Perrin, O.~McCabe, G.~Everly, and J.~Links.
\newblock Preparing for an influenza pandemic: Mental health considerations.
\newblock {\em Prehospital and Disaster Medicine}, 24(3), 2009.

\bibitem{pip_1}
O.~M. Araz, P.~Damien, D.~A. Paltiel, S.~Burke, B.~van~de Geijn, A.~Galvani, and L.~A. MEyers.
\newblock Simulating school closure policies for cost effective pandemic decision making.
\newblock {\em BMC Public Health}, page 449, 2012.

\bibitem{base_paper}
A.~Alexi, A.~Rosenfeld, and T.~Lazebnik.
\newblock Multi-species prey–predator dynamics during a multi-strain pandemic.
\newblock {\em Chaos: An Interdisciplinary Journal of Nonlinear Science}, 33(7):073106, 2023.

\bibitem{rabis_israel}
R.~King, M.~Eyngor, S.~Novak, M.~P. Markovich, T.~Goshen, N.~Edery, R.~Lapid, A.~Reichman, J.~L. Maki, E.~W. Lankau, and B.~Yakobson.
\newblock Oral vaccination and population management focused on juvenile golden jackals halts a rabies epizootic in israel.
\newblock {\em Israel Journal of Veterinary Medicine}, 79(1), 2024.

\bibitem{rabis_bad}
C.~Fisher, D.~Streicker, and M.~Schnell.
\newblock The spread and evolution of rabies virus: conquering new frontiers.
\newblock {\em Nat Rev Microbiol}, 16:241--255, 2018.

\bibitem{orr_r_2}
E.~Frank and A.~Sudarshan.
\newblock The social costs of keystone species collapse: Evidence from the decline of vultures in india.
\newblock {\em American Economic Review}, 114(10):3007--3040, 2024.

\bibitem{rabis_bite_spread}
D.~G. Hankins and J.~A. Rosekrans.
\newblock Overview, prevention, and treatment of rabies.
\newblock {\em Mayo Clinic Proceedings}, 79(5):671--676, 2004.

\bibitem{rabis_jackle_1}
J.~Bingham, C.~M. Foggin, A.~I. Wandeler, and F.~W.~G. Hill.
\newblock The epidemiology of rabies in zimbabwe. 2. rabies in jackals ({\textit{canis adustus}} and {\textit{canis mesomelas}}).
\newblock {\em Onderstepoort J Vet Res}, 66:11--23, 1999.

\bibitem{rabis_jackle_2}
F.~Cliquet, J.~P. Gurbuxani, H.~K. Pradhan, B.~Pattnaik, S.~S. Patil, A.~Regnault, H.~Begouen, A.~L. Guiot, R.~Sood, P.~Mahl, R.~Singh, F.~X. Meslin, E.~Picard, M.~F.~A. Aubert, and J.~Barrat.
\newblock The safety and efficacy of the oral rabies vaccine sag2 in indian stray dogs.
\newblock {\em Vaccine}, 25(17):3409--3418, 2007.

\bibitem{rabits_g_1}
S.~P.~D. Riley, J.~Hadidian, and D.~A. Manski.
\newblock Population density, survival, and rabies in raccoons in an urban national park.
\newblock {\em Canadian Journal of Zoology}, 76(6):1153--1164, 1998.

\bibitem{rabits_g_2}
R.~M. Anderson, H.~C. Jackson, R.~M. May, and A.~M. Smith.
\newblock Population dynamics of fox rabies in europe.
\newblock {\em Nature}, 289:765--771, 1981.

\bibitem{rabits_g_3}
A.~C. Banyard, D.~T.~S. Hayman, C.~M. Freuling, T.~M{\"u}ller, A.~R. Fooks, and N.~Johnson.
\newblock Chapter 6 - bat rabies.
\newblock In {\em Rabies (Third Edition): Scientific Basis of the Disease and Its Management}, pages 215--267. 2013.

\bibitem{rabis_inside_host}
G.~Ugolini.
\newblock Chapter 10 - rabies virus as a transneuronal tracer of neuronal connections.
\newblock In A.~C. Jackson, editor, {\em Research Advances in Rabies}, volume~79 of {\em Advances in Virus Research}, pages 165--202. Academic Press, 2011.

\bibitem{rabis_fatal}
K.~C. Bloch.
\newblock Rabies: Still a uniformly fatal disease? historical occurrence, epidemiological trends, and paradigm shifts.
\newblock {\em Current Infectious Disease Reports}, 14(4):408--422, 2012.
\newblock Central Nervous System and Eye Infections (KC Bloch, Section Editor).

\bibitem{Rupprecht2024}
C.~E. Rupprecht, T.~Buchanan, F.~Cliquet, R.~King, and T.~Müller.
\newblock A global perspective on oral vaccination of wildlife against rabies.
\newblock {\em Journal of Wildlife Diseases}, 60(2):241--284, 2024.

\bibitem{morters_control_rabies_2013}
M.~K. Morters, O.~Restif, K.~Hampson, S.~Cleaveland, J.~L.~N. Wood, and A.~J.~K. Conlan.
\newblock Evidence-based control of canine rabies: A critical review of population density reduction.
\newblock {\em Journal of Animal Ecology}, 82(1):6--14, 2013.

\bibitem{gdalevich2000rabies}
M.~Gdalevich, D.~Mimouni, I.~Ashkenazi, and J.~Shemer.
\newblock Rabies in israel: decades of prevention and a human case.
\newblock {\em Public Health}, 114(6):484--487, 2000.

\bibitem{Fretwell1970}
S.~D. Fretwell and J.~H.~L. Lucas.
\newblock On territorial behavior and other factors influencing habitat distribution in birds: {I}. theoretical development.
\newblock {\em Acta Biotheor.}, 19:16--36, 1970.

\bibitem{social_dist}
S.~Stockmaier, N.~Stroeymeyt, E.~C. Shattuck, D.~M. Hawley, L~A. Meyers, and D.~I. Bolnick.
\newblock Infectious diseases and social distancing in nature.
\newblock {\em Science}, 371(6533):TBA, 2021.

\bibitem{sick_walk_2}
K.~Hueffer, S.~Khatri, S.~Rideout, M.~B. Haris, R.~L. Papke, C.~Stokes, and M.~K. Schulte.
\newblock Rabies virus modifies host behaviour through a snake-toxin like region of its glycoprotein that inhibits neurotransmitter receptors in the cns.
\newblock {\em Sci Rep}, 2017.

\bibitem{no_recover_class}
J.~E. Childs, A.~T. Curns, M.~E. Dey, L.~A. Real, L.~Feinstein, and O.~N. Bjornstad.
\newblock Predicting the local dynamics of epizootic rabies among raccoons in the united states.
\newblock {\em PNAS}, 97:13666--13671, 2000.

\bibitem{orr_cool_paper}
B.~A. Lerch and C.~A. Karen.
\newblock A flexible theory for the dynamics of social populations: Within-group density dependence and between-group processes.
\newblock {\em Ecological Monographs}, 94(2):e1604, 2024.

\bibitem{DiQuinzio2008}
M.~Di~Quinzio and A.~McCarthy.
\newblock Rabies risk among travellers.
\newblock {\em CMAJ: Canadian Medical Association Journal}, 178(5):567, 2008.

\bibitem{Castello2018}
J.~R. Castelló.
\newblock {\em Canids of the World: Wolves, Wild Dogs, Foxes, Jackals, Coyotes, and Their Relatives}.
\newblock Princeton University Press, Princeton, NJ, 2018.

\bibitem{exposed_duration}
V.~de~Vos.
\newblock Pseudopregnancy in the black-backed jackal (chain mesomelas schreber).
\newblock {\em Jl S. Afr. Vet. Med. Ass.}, 40(4):381--383, 1970.

\bibitem{Lange2021}
P.N.A.M.J.G. Lange, G.~Lelieveld, and H.J. De~Knegt.
\newblock Diet composition of the golden jackal \textit{Canis aureus} in south-east europe – a review.
\newblock {\em Mammal Review}, 51:207--213, 2021.

\bibitem{vaccine_duration}
J.~Maki, A.~L. Guiot, M.~Aubert, B.~Brochier, F.~Cliquet, C.~A. Hanlon, R.~King, E.~H. Oertli, C.~E. Rupprecht, C.~Schumacher, D.~Slate, B.~Yakobson, A.~Wohlers, and E.~W. Lankau.
\newblock Oral vaccination of wildlife using a vaccinia-rabies-glycoprotein recombinant virus vaccine (raboral v-rg®): a global review.
\newblock {\em Vet Res}, 22(48):107221, 2017.

\bibitem{teddy_sir_review}
T.~Lazebnik.
\newblock Computational applications of extended sir models: A review focused on airborne pandemics.
\newblock {\em Ecological Modelling}, 483:110422, 2023.

\bibitem{metric_paper_3}
D.~Breda, F.~Florian, J.~Ripoll, and R.~Vermiglio.
\newblock Efficient numerical computation of the basic reproduction number for structured populations.
\newblock {\em Journal of Computational and Applied Mathematics}, 384:113165, 2021.

\bibitem{rosenfeld}
A.~Rosenfeld, O.~Maksomov, and S.~Kraus.
\newblock When security games hit traffic: A deployed optimal traffic enforcement system.
\newblock {\em Artificial Intelligence}, 289:103381, 2020.

\bibitem{alexi_games}
A.~Alexi, A.~Rosenfeld, and T.~Lazebnik.
\newblock A security games inspired approach for distributed control of pandemic spread.
\newblock {\em Advanced Theory and Simulations}, 6(2):2200631, 2023.

\bibitem{norm_to_graph_based}
T.~Lazebnik.
\newblock Transforming norm-based to graph-based spatial representation for spatio-temporal epidemiological models.
\newblock {\em arXiv}, 2024.

\bibitem{tskm_intro}
B.~Cai, G.~Huang, N.~Samadiani, G.~Li, and C-H. Chi.
\newblock Efficient time series clustering by minimizing dynamic time warping utilization.
\newblock {\em IEEE Access}, 9:46589--46599, 2021.

\bibitem{autoencoder}
D.~Bank, N.~Koenigstein, and R.~Giryes.
\newblock {\em Autoencoders}, pages 353--374.
\newblock Springer International Publishing, Cham, 2023.

\bibitem{teddy_cancer}
T.~Lazebnik and L.~Simon-Keren.
\newblock Cancer-inspired genomics mapper model for the generation of synthetic dna sequences with desired genomics signatures.
\newblock {\em Computers in Biology and Medicine}, 164:107221, 2023.

\bibitem{metric_paper_1}
L.~Di~Domenico, G.~Pullano, C.~E. Sabbatini, P.~Y. Bo~Elle, and V.~Colizza.
\newblock Impact of lockdown on covid-19 epidemic in ile-de-france and possible exit strategies.
\newblock {\em BMC Medicine}, 2020.

\bibitem{metric_paper_2}
S.~Zhao, L.~Stone, D.~Gao, S.~S. Musa, M.~K.~C. Chong, D.~He, and M.~H. Wang.
\newblock Imitation dynamics in the mitigation of the novel coronavirus disease (covid-19) outbreak in wuhan, china from 2019 to 2020.
\newblock {\em Annals of Transnational Medicine}, 8, 2020.

\bibitem{metric_paper_4}
K.~Chatterjee, K.~Chatterjee, A.~Kumar, and S.~Shankar.
\newblock Healthcare impact of covid-19 epidemic in india: A stochastic mathematical model.
\newblock {\em Medical Journal Armed Forces India}, 76(2):147--155, 2020.

\bibitem{appendix}
R.~P. Agarwal and V.~Lakshmikantham.
\newblock Uniqueness and nonuniqueness criteria for ordinary differential equations.
\newblock {\em World Scientific}, 1993.

\bibitem{sveta_appendix}
S.~Bunimovich-Mendrazitsky, J.~C. {Gluckman}, and J.~Chaskalovic.
\newblock A mathematical model of combined bacillus calmette-guerin (bcg) and interleukin (il)-2 immunotherapy of superficial bladder cancer.
\newblock {\em Journal of Theoretical Biology}, 277(1):27--40, 2011.

\bibitem{final_appendix}
M.~Schatzman.
\newblock Numerical analysis: a mathematical introduction.
\newblock {\em Oxford Univ. Press.}, 2002.

\bibitem{ngm}
O.~Diekmann, J.~A.~P. Heesterbeek, and M.~G. Roberts.
\newblock The construction of next-generation matrices for compartmental epidemic models.
\newblock {\em Journal of the Royal Society interface}, 164:107221, 2010.

\bibitem{doCoutoReis2015}
R.~do~Couto~Reis.
\newblock {\em Distribution patterns and genetic structure of golden jackal in Europe and Asia}.
\newblock Phd thesis, University of Porto, 2015.

\bibitem{Rotem2011}
G.~Rotem, H.~Berger, R.~King, P.~B. Kutiel, and D.~Saltz.
\newblock The effect of anthropogenic resources on the space-use patterns of golden jackals.
\newblock {\em The Journal of Wildlife Management}, 75:132--136, 2011.

\bibitem{Beardsworth2022}
C.~E. Beardsworth, E.~Gobbens, F.~van Maarseveen, B.~Denissen, A.~Dekinga, R.~Nathan, S.~Toledo, and A.~I. Bijleveld.
\newblock Validating atlas: A regional-scale high-throughput tracking system.
\newblock {\em Methods in Ecology and Evolution}, 13:1990--2004, 2022.

\bibitem{arnon_wildlife_tracking_2023}
E.~Arnon, S.~Cain, A.~Uzan, R.~Nathan, O.~Spiegel, and S.~Toledo.
\newblock Wildlife tracking.
\newblock {\em Sensors}, pages 1--23, 2023.

\bibitem{jw1}
Y.~Kebede.
\newblock A review on: Distribution, ecology and status of golden jackal (canis aureus) in africa.
\newblock {\em Journal of Natural Sciences Research}, 7(1):32--43, 2017.

\bibitem{jw2}
R.~P.~D. Atkinson, C.~J. Rhodes, D.~W. Macdonald, and R.~M. Anderson.
\newblock Scale-free dynamics in the movement patterns of jackals.
\newblock {\em Oikos}, 98(1):134--140, 2002.

\bibitem{orr_r_4}
S.~Cain, T.~Solomon, Y.~Leshem, S.~Toledo, E.~Arnon, A.~Roulin, and O.~Spiegel.
\newblock Movement predictability of individual barn owls facilitates estimation of home range size and survival.
\newblock {\em Movement Ecology}, 11(1):1--13, 2023.

\bibitem{weather_api_2}
A.~Musah, L.~M.~M. Dutra, A.~Aldosery, E.~Browning, T.~Ambrizzi, I.~V.~G. Borges, M.~Tunali, S.~Basibuyuk, O.~Yenigun, G.~M.~M. Moreno, A.~C.~G. da~Silva, W.~P. dos Santos, C.~L. de~Lima, T.~Massoni, K.~E. Jones, L.~C. Campos, and P.~Kostkova.
\newblock An evaluation of the openweathermap api versus inmet using weather data from two brazilian cities: Recife and campina grande.
\newblock {\em Data}, 7(8), 2022.

\bibitem{weather_api_1}
C.~Dewi and R-C. Chen.
\newblock Integrating real-time weather forecasts data using openweathermap and twitter.
\newblock {\em International Journal of Information Technology and Business}, 1, 2019.

\bibitem{mcknight2010mann}
P.~E. McKnight and J.~Najab.
\newblock Mann-whitney u test.
\newblock {\em The Corsini encyclopedia of psychology}, pages 1--1, 2010.

\bibitem{teddy_economy}
T.~Lazebnik, L.~Shami, and S.~Bunimovich-Mendrazitsky.
\newblock Spatio-temporal influence of non-pharmaceutical interventions policies on pandemic dynamics and the economy: The case of covid-19.
\newblock {\em Research Economics}, 2021.

\bibitem{jones2021spread}
A.~Jones and N.~Strigul.
\newblock Is spread of covid-19 a chaotic epidemic?
\newblock {\em Chaos, Solitons \& Fractals}, 142:110376, 2021.

\bibitem{sene2020sir}
N.~Sene.
\newblock Sir epidemic model with mittag--leffler fractional derivative.
\newblock {\em Chaos, Solitons \& Fractals}, 137:109833, 2020.

\bibitem{ahmed2023bifurcation}
M.~Ahmed, A.~B. Masud, and M.~A. Sarker.
\newblock Bifurcation analysis and optimal control of discrete sir model for covid-19.
\newblock {\em Chaos, Solitons \& Fractals}, 174:113899, 2023.

\bibitem{Dougherty2022}
E.~R. Dougherty, D.~P. Seidel, J.~K. Blackburn, W.~C. Turner, and W.~M. Getz.
\newblock A framework for integrating inferred movement behavior into disease risk models.
\newblock {\em Movement Ecology}, 10:31, 2022.

\bibitem{Dobson1986}
A.~P. Dobson and P.~J. Hudson.
\newblock Parasites, disease and the structure of ecological communities.
\newblock {\em Trends in Ecology and Evolution}, 1(1):11--15, 1986.

\bibitem{Daszak2000}
P.~Daszak, A.~A. Cunningham, and A.~D. Hyatt.
\newblock Emerging infectious diseases of wildlife: Threats to biodiversity and human health.
\newblock {\em Science}, 287(5452):443--449, 2000.

\bibitem{Frank2024}
E.~Frank and A.~Sudarshan.
\newblock The social costs of keystone species collapse: Evidence from the decline of vultures in india.
\newblock {\em American Economic Review}, 114(10):3007--3040, 2024.

\bibitem{Spiegel2022}
O.~Spiegel, N.~Anglister, and M.~M. Crafton.
\newblock Movement data provides insight into feedbacks and heterogeneities in host--parasite interactions.
\newblock In V.~Ezenwa, S.~M. Altizer, and R.~Hall, editors, {\em Animal Behavior and Parasitism}. 2022.

\bibitem{Anglister2024}
N.~Anglister, S.~Gonen-Shalom, P.~Shlanger, E.~Blotnick-Rubin, A.~Rosenzweig, I.~Horowitz, O.~Hatzofe, R.~King, L.~Anglister, and O.~Spiegel.
\newblock Plasma cholinesterase activity: A benchmark for rapid detection of pesticide poisoning in an avian scavenger.
\newblock {\em Science of the Total Environment}, 877:162903, 2024.

\bibitem{Ezenwa2016}
V.~O. Ezenwa, E.~A. Archie, M.~E. Craft, D.~M. Hawley, L.~B. Martin, J.~Moore, and L.~White.
\newblock Host behaviour-parasite feedback: an essential link between animal behaviour and disease ecology.
\newblock {\em Proceedings. Biological Sciences}, 283(1828):20153078, 2016.

\bibitem{wong2020spreading}
D.~W.~S. Wong and Y.~Li.
\newblock Spreading of covid-19: Density matters.
\newblock {\em Plos one}, 15(12):e0242398, 2020.

\bibitem{moreno2023stocking}
M.~J. Moreno-Madri{\~n}an and E.~Kontowicz.
\newblock Stocking density and homogeneity, considerations on pandemic potential.
\newblock {\em Zoonotic Diseases}, 3(2):85--92, 2023.

\bibitem{yin2021association}
H.~Yin, T.~Sun, L.~Yao, Y.~Jiao, L.~Ma, L.~Lin, J.~C. Graff, L.~Aleya, A.~Postlethwaite, W.~Gu, et~al.
\newblock Association between population density and infection rate suggests the importance of social distancing and travel restriction in reducing the covid-19 pandemic.
\newblock {\em Environmental Science and Pollution Research}, pages 1--7, 2021.

\bibitem{Morters2013}
M.~K. Morters, O.~Restif, K.~Hampson, S.~Cleaveland, J.~L. Wood, and A.~J. Conlan.
\newblock Evidence-based control of canine rabies: a critical review of population density reduction.
\newblock {\em The Journal of Animal Ecology}, 82(1):7--14, 2013.

\bibitem{moorcroft2006mechanistic}
P.~R. Moorcroft, M.~A. Lewis, and R.~L. Crabtree.
\newblock Mechanistic home range models capture spatial patterns and dynamics of coyote territories in yellowstone.
\newblock {\em Proceedings of the Royal Society B}, 273(1594):1651--1659, 2006.

\bibitem{potts2012territorial}
J.~R. Potts, S.~Harris, and L.~Giuggioli.
\newblock Territorial dynamics and stable home range formation for central place foragers.
\newblock {\em PloS One}, 7(3):e34033, 2012.

\bibitem{rozins2018social}
C.~Rozins, M.~J. Silk, D.~P. Croft, R.~J. Delahay, D.~J. Hodgson, R.~A. McDonald, N.~Weber, and M.~Boots.
\newblock Social structure contains epidemics and regulates individual roles in disease transmission in a group-living mammal.
\newblock {\em Ecology and Evolution}, 8(23):12044--12055, 2018.

\bibitem{viana2023effects}
M.~Viana, J.~A. Benavides, A.~Broos, D.~Iba{\~n}ez~Loayza, R.~Ni{\~n}o, J.~Bone, A.~da~Silva~Filipe, R.~Orton, W.~Valderrama~Bazan, J.~Matthiopoulos, and D.~G. Streicker.
\newblock Effects of culling vampire bats on the spatial spread and spillover of rabies virus.
\newblock {\em Science Advances}, 9(10):eadd7437, 2023.

\bibitem{downing2023culling}
B.~C. Downing, M.~J. Silk, R.~J. Delahay, S.~Bearhop, and N.~J. Royle.
\newblock Culling-induced perturbation of social networks of wild geese reinforces rather than disrupts associations among survivors.
\newblock {\em Journal of Applied Ecology}, 60:2613--2624, 2023.

\bibitem{cooper}
I.~Cooper, A.~Mondal, and C.~G. Antonopoulos.
\newblock A {SIR} model assumption for the spread of covid-19 in different communities.
\newblock {\em Chaos, Solitons and Fractals}, 139:110057, 2020.

\bibitem{sir_example_2}
W.~Chen.
\newblock A mathematical model of ebola virus based on sir model.
\newblock In {\em 2015 International Conference on Industrial Informatics - Computing Technology, Intelligent Technology, Industrial Information Integration}, pages 213--216, 2015.

\bibitem{dang}
Y-X. Dang, X-Z. Li, and M.~Martcheva.
\newblock Competitive exclusion in a multi-strain immuno-epidemiological influenza model with environmental transmission.
\newblock {\em Journal of Biological Dynamics}, 10(1), 2016.

\bibitem{levy2015modeling}
A.~Levy.
\newblock Modeling without models.
\newblock {\em Philosophical Studies}, 172:781--798, 2015.

\bibitem{spiegel2017s}
O.~Spiegel, S.~T. Leu, C.~M. Bull, and A.~Sih.
\newblock What's your move? movement as a link between personality and spatial dynamics in animal populations.
\newblock {\em Ecology letters}, 20(1):3--18, 2017.

\bibitem{teddy_cows}
T.~Lazebnik and O.~Spiegel.
\newblock Individual variation affects outbreak magnitude and predictability in an extended multi-pathogen sir model of pigeons vising dairy farms.
\newblock {\em Ecological modeling}, 2024.

\bibitem{large_models_not_fit}
V.~Vytla, S.~K. Ramakuri, K.~K. Peddi, A.~Srinivas, and N.~N. Ragav.
\newblock Mathematical models for predicting covid-19 pandemic: A review.
\newblock {\em Journal of Physics: Conference Series}, 1797:012009, 2021.

\bibitem{teddy_ariel}
T.~Lazebnik and A.~Alexi.
\newblock Comparison of pandemic intervention policies in several building types using heterogeneous population model.
\newblock {\em Communications in Nonlinear Science and Numerical Simulation}, 107(4):106176, 2022.

\bibitem{vsprem2024factors}
N.~{\v{S}}prem, V.~Baruk{\v{c}}i{\'c}, A.~Jazbec, D.~Ugarkovi{\'c}, I.~Ili{\'c}, and B.~Pokorny.
\newblock Factors affecting hunting efficiency in the case of golden jackal.
\newblock {\em European journal of wildlife research}, 70(2):19, 2024.

\bibitem{or_sir_example}
P.~Sah, S.~T. Leu, P.~C. Cross, P.~J. Hudson, and S.~Bansal.
\newblock Unraveling the disease consequences and mechanisms of modular structure in animal social networks.
\newblock {\em Proc Natl Acad Sci U S A}, 18(114):4165--4170, 2017.

\bibitem{con_new_2}
E.~F. Stuber, B.~S. Carlson, and B.~R. Jesmer.
\newblock {Spatial personalities: a meta-analysis of consistent individual differences in spatial behavior}.
\newblock {\em Behavioral Ecology}, 33(3):477--486, 2022.

\bibitem{lazebnik2024exploration}
T.~Lazebnik, Y.~Golov, R.~Gurka, A.~Harari, and A.~Liberzon.
\newblock Exploration--exploitation model of moth-inspired olfactory navigation.
\newblock {\em Journal of the Royal Society Interface}, 21(216):20230746, 2024.

\bibitem{eve_nature_1}
K.~Mehlhorn, B.~R. Newell, P.~M. Todd, M.~D. Lee, K.~Morgan, V.~A. Braithwaite, D.~Hausmann, K.~Fiedler, and C.~Gonzalez.
\newblock Unpacking the exploration--exploitation tradeoff: A synthesis of human and animal literatures.
\newblock {\em Decision}, 2(3):191--215, 2015.

\bibitem{eve_nature_4}
C.~T. Monk, M.~Barbier, P.~Romanczuk, J.~R. Watson, J.~Alos, S.~Nakayama, D.~I. Rubenstein, S.~A. Levin, and R.~Arlinghaus.
\newblock How ecology shapes exploitation: a framework to predict the behavioural response of human and animal foragers along exploration–exploitation trade-offs.
\newblock {\em Ecology Letters}, 21(6):779--793, 2018.

\bibitem{eve_nature_5}
F.~Cinotti, V.~Fresno, N.~Aklil, E.~Coutureau, B.~Girard, A.~R. Marchand, and M.~Khamassi.
\newblock Dopamine blockade impairs the exploration-exploitation trade-off in rats.
\newblock {\em Scientific Reports}, 9(1):6770, 2019.

\bibitem{SPIEGEL201690}
O.~Spiegel and M.~C. Crofoot.
\newblock The feedback between where we go and what we know—information shapes movement, but movement also impacts information acquisition.
\newblock {\em Current Opinion in Behavioral Sciences}, 12:90--96, 2016.

\bibitem{con_new_6}
P.~Trinh, J.~R. Zaneveld, S.~Safranek, and P.~M. Rabinowitz.
\newblock One health relationships between human, animal, and environmental microbiomes: A mini-review.
\newblock {\em Frontiers in Public Health}, 6, 2018.

\bibitem{orr_end_1}
R.~Nathan, Christopher~T. Monk, R.~Arlinghaus, T.~Adam, J.~Alos, M.~Assaf, H.~Baktoft, C.~E. Beardsworth, M.~G. Bertram, A.~I. Bijleveld, T.~Brodin, J.~L. Brooks, A.~Campos-Candela, S.~J. Cooke, K.~O. Gjelland, P.~R. Gupte, R.~Harel, G.~Hellström, F.~Jeltsch, S.~S. Killen, T.~Klefoth, R.~Langrock, R.~J. Lennox, E.~Lourie, J.~R. Madden, Y.~Orchan, I.~S. Pauwels, M.~Riha, M.~Roeleke, U.~E. Schlagel, D.~Shohami, J.~Signer, S.~Toledo, O.~Vilk, S.~Westrelin, M.~A. Whiteside, and I.~Jaric.
\newblock Big-data approaches lead to an increased understanding of the ecology of animal movement.
\newblock {\em Science}, 375(6582), 2022.

\bibitem{Barrile2024}
G.~M. Barrile, P.~C. Cross, C.~Stewart, J.~Malmberg, R.~P. Jakopak, J.~Binfet, K.~L. Monteith, B.~Werner, J.~Jennings-Gaines, and J.~A. Merkle.
\newblock Chronic wasting disease alters the movement behavior and habitat use of mule deer during clinical stages of infection.
\newblock {\em Ecology and Evolution}, 14:e11418, 2024.

\bibitem{Grabow2024}
M.~Grabow, W.~Ullmann, C.~Landgraf, R.~Sollmann, C.~Scholz, R.~Nathan, S.~Toledo, R.~Luhken, J.~Fickel, F.~Jeltsch, N.~Blaum, V.~Radchuk, R.~Tiedemann, and S.~Kramer-Schadt.
\newblock Sick without signs. subclinical infections reduce local movements, alter habitat selection, and cause demographic shifts.
\newblock {\em Communications Biology}, 7(1):1426, 2024.

\end{thebibliography}
\bibliographystyle{unsrt}

\section*{Appendix}
\begin{figure}
    \centering
    \includegraphics[width=0.8\textwidth]{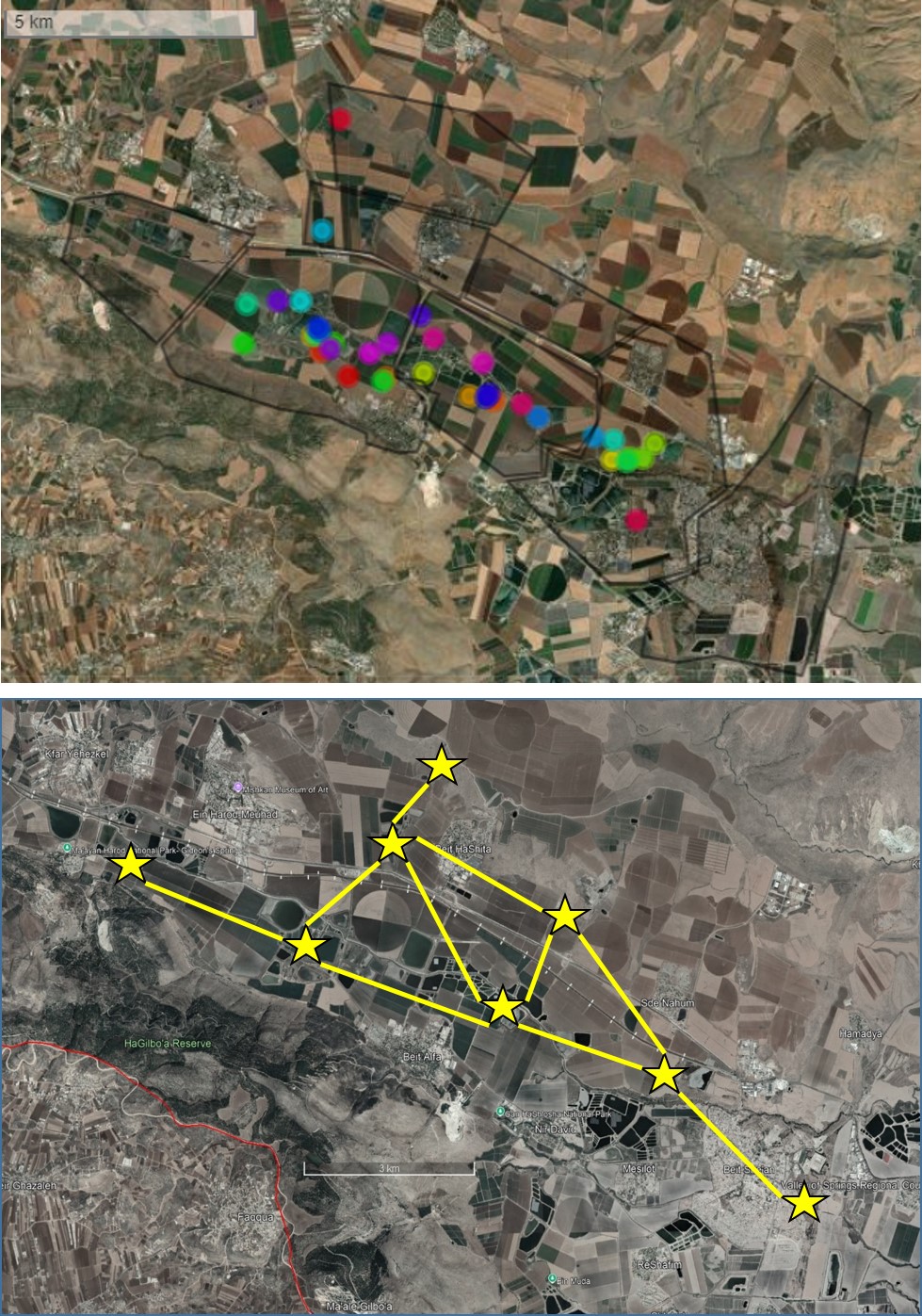}
    \caption{A map of the study area. The upper panel shows the centroid of each jackal (different colors), and the inferred activity centers (black line polygons). The lower panel shows the mathematical graph of connections among activity centers (yellow stars).}
    \label{fig:map}
\end{figure}

Fig. \ref{fig:map} shows the activity centers of different jackal groups as depicted in the collected dataset from the Harod and HaMaayanot valleys in the North-East of Israel. Each color indicates a centroid of an ATLAS-tracked individual. The black-colored polygons indicate heuristically the activity centers used for our simulation which are based on both the raw movement data of the jackals and experts' familiarity with the area. In general, jackals were grouped into an activity center if they tended to move together, and had close-by centroids. Additional considerations were geographical barriers (e.g. the highway), jackal's activity monitored with camera traps (e.g. in activity centers without sufficient tracking coverage). We note that while the suggested activity centers are obviously simplistic and crude (e.g. including straight lines and ignoring some biological features), yet, for our model the actual shape and boundaries of these centers are insignificant, because the model is based on the mathematical graph of connectivity among activity centers.   

\begin{figure}[!ht]
    \centering
    \includegraphics[width=0.99\textwidth]{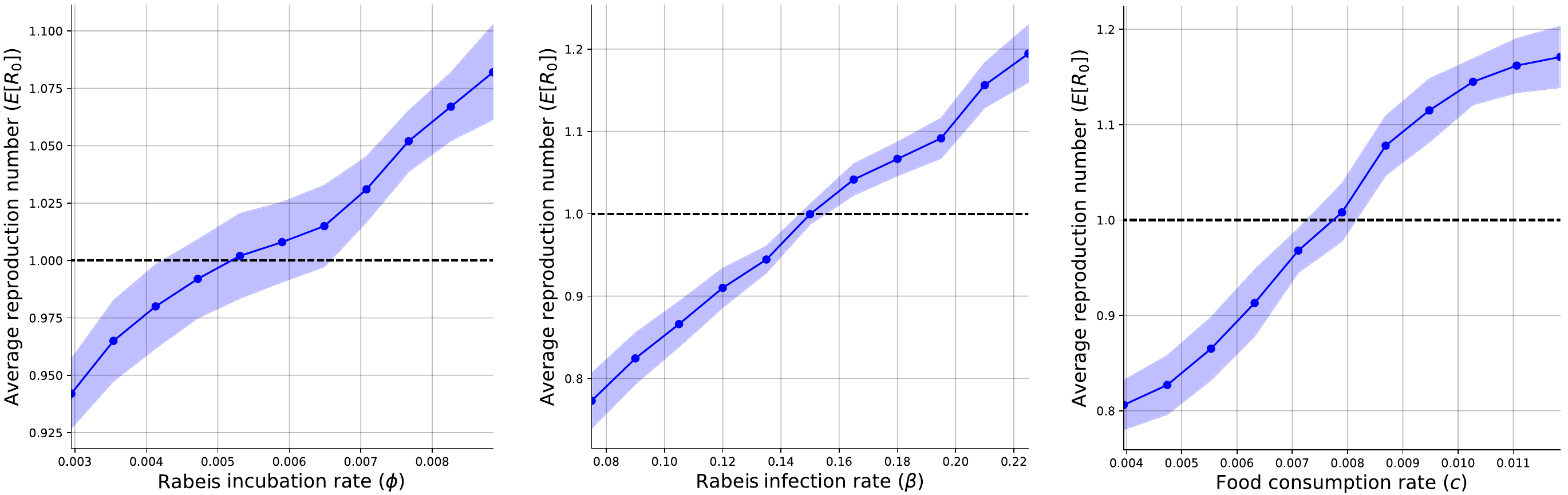}
\caption{A sensitivity analysis of the jackal-related properties. The results are shown as the mean \(\pm\) standard deviation of \(n=100\) repetitions.} 
\label{fig:sensitivity}
\end{figure}

In order to assess the robustness of the results to the jackal-related physical and behavioral properties, we conducted a one-dimensional sensitivity analysis with the ARN for the case of no EIPs. Fig. \ref{fig:sensitivity} presents a sensitivity analysis of the jackal-related properties in terms of the ARN. The results are shown as the mean \(\pm\) standard deviation of \(n=100\) repetitions. The x-axis range is taken to be between 50\% and 150\% of the value provided in Table \ref{table:params}. Specifically, the left shows a monotonically increasing relationship between the rabies' incubation rate (\(\psi\)) and the ARN. Similarly, the central panel presents a monotonically increasing relationship between the rabies infection rate (\(\beta\)) and the ARN. Moreover, the right-hand panel describes a monotonically increasing relationship between the food consumption rate (\(c\)) and the ARN. 

\end{document}